\documentclass[a4paper,11pt]{article}

\usepackage[top=2.5cm,bottom=2.5cm,left=2cm,right=2cm]{geometry} 
\usepackage{indentfirst}

\usepackage{amsmath, amssymb} 
\if
\makeatletter
 
 \@addtoreset{equation}{section}
\makeatother
\fi

\usepackage{graphicx} 
\usepackage{float}
\graphicspath{{./figure/}} 
\usepackage[labelsep=period,figurename=Fig. , tablename=Table]{caption} 

\usepackage{booktabs}
\usepackage{multirow}

\usepackage{url} 

\usepackage{bm} 

\title{Statistical properties of position-dependent
ball-passing networks\\ in football games}

\usepackage{authblk}
\author[1]{Takuma Narizuka\thanks{Corresponding author. Department of Physics, School of Advanced Science and Engineering, Waseda University, Shinjuku, Tokyo 169-8555, Japan. Tel: +81 3 5286 8187.\\
{\it Email address}: physicist.t.n@fuji.waseda.jp (T. Narizuka).\\
}}
\author[2]{Ken Yamamoto}
\author[1]{Yoshihiro Yamazaki}
\affil[1]{Department of Physics, School of Advanced Science and Engineering, Waseda University, Shinjuku, Tokyo 169-8555, Japan}
\affil[2]{Department of Physics, Faculty of Science and Engineering, Chuo University, Bunkyo, Tokyo 112-8551, Japan}

\date{}

\begin{document}
	\maketitle
\begin{abstract}
Statistical properties of position-dependent ball-passing networks in real football games are examined. We find that the networks have the small-world property, and their degree distributions are fitted well by a truncated gamma distribution function. In order to reproduce these properties of networks, a model based on a Markov chain is proposed. 
\end{abstract}

\baselineskip 14pt
\renewcommand{\arraystretch}{1.2}
\section{Introduction}
Scientific studies on sports activities have been carried out in a wide variety of research fields such as psychology, physiology, biomechanics, and also physics.
Among various sports events, football is one of the major subjects \cite{John}.
From the viewpoint of physics, studies with respect to football games are considered to be classified into the following two types: mechanical and statistical.
In the latter case, goal distribution \cite{Reep1971,Malacarne2000a,Greenhough2002,Bittner2008} and outcome prediction of football games \cite{Ben-Naim2007a,Ribeiro2010,Heuer2009,Heuer2010}
have been studied for example.
A football game can be considered as a dynamical system in which game players interact with each other via one ball.
Ball-passing events have been focused on
in various statistical analyses for 
the collective behavior of players \cite{Yue2008a,Yue2008b},
temporal sequences of players' action \cite{Borrie2002}
and ball movements\cite{Mendes2007a}, and
passing sequence to goal \cite{Hughes2005}.

In statistical physics, analysis for complex networks has achieved rapid development recently \cite{Watts1998, Barabasi1999a}.
The network analysis has already been applied to football games such as
a structural property of ball-passing networks \cite{Yamamoto2010,Yamamoto2011b}, and assessment of players \cite{Duch2010,Javier2012}.
In the studies of the ball-passing networks \cite{Yamamoto2010,Yamamoto2011b}, each node and edge of the network represent an individual player and passing of the ball, respectively.
One main conclusion in their studies is that ball-passing networks of football games have the scale-free property, namely the degree distributions follow the power law.
However, in their network analysis, the total number of nodes were only 11, which was the number of players on a ground in one team.
Clearly, it is too few to judge the power-law behavior of the degree distributions.

In this paper we propose another method for creating a ball-passing network and report statistical properties of the network.
Since each player has his own role corresponding to their home position and it is important factor for ball passing, we create a {\it position-dependent} network in the next section.
In section 3, the structural properties and the degree distributions of the networks obtained from real games are examined.
We find that the degree distributions can be fitted with
a truncated gamma distribution in common.
In section 4, we propose a numerical model based on a Markov chain
by introducing the ball-possession probability.
Discussion and conclusion are given in sections 5 and 6, respectively.

\section{Method for creating a ball-passing network}
\begin{figure}[H]
	\centering
		\includegraphics[width=9cm]{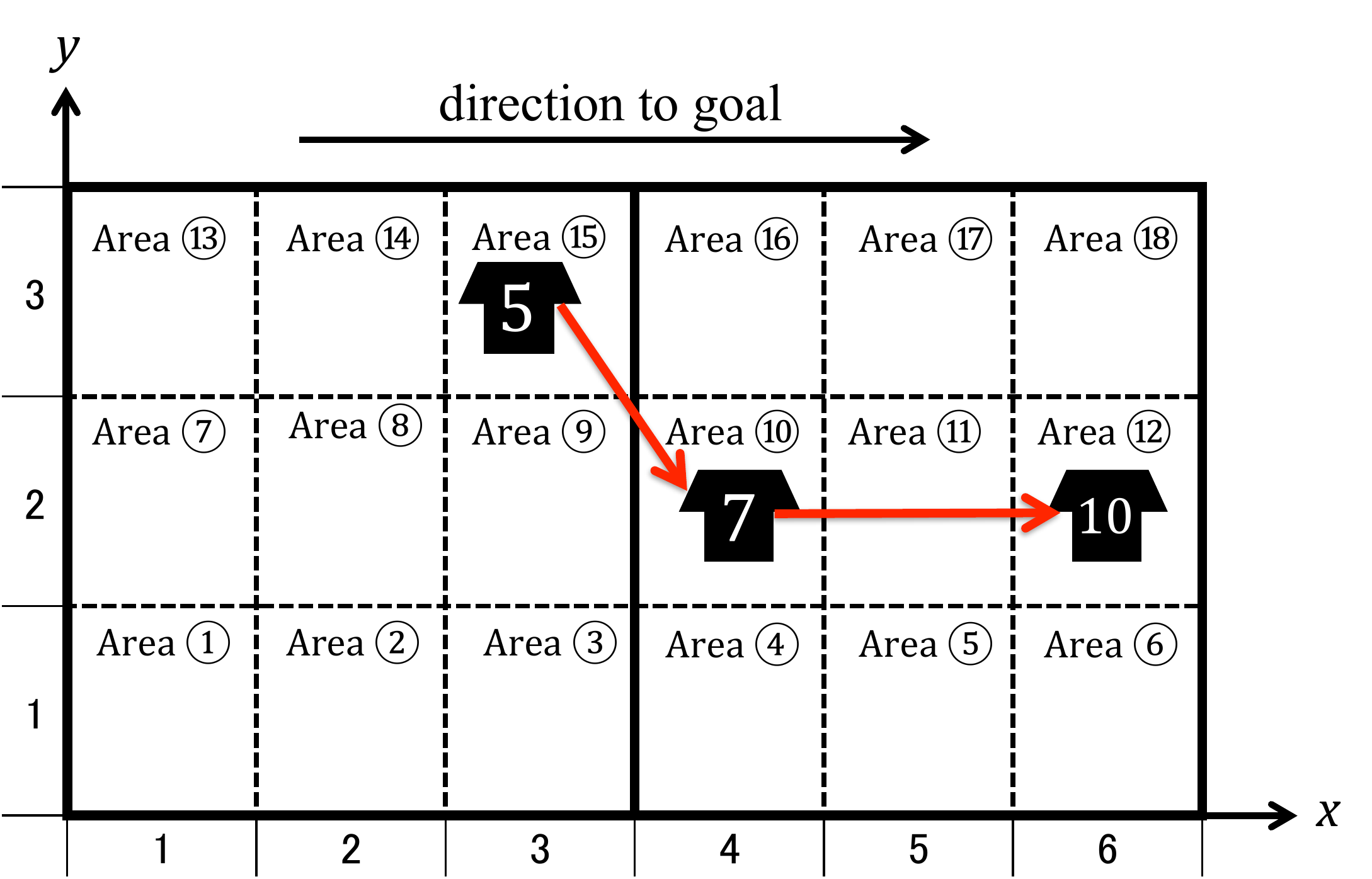}
		\caption{Soccer field divided into 18 areas.}
		\label{fig:field}
\end{figure}

A ``position-dependent'' network of ball passing,
where the position of a player in a soccer field is considered,
was obtained by the following method.
In order to specify the position of a player,
we divide the field into 18 areas 
(six areas along the goal direction, and three areas along the vertical direction in Fig. \ref{fig:field}).
Note that this is the same division as used in the FIFA official statistical data of 2010 World Cup South Africa \cite{FIFA} and in the reference \cite{Borrie2002}.
An area expressed in the coordinate $(x , y)$ is labeled by the area number $A_{xy} = 6(y-1)+x$ ($1\leq x\leq6$ , $1\leq y\leq3$). 
A node is assigned to a player on one of the 18 areas, and labeled by the node number $11(A_{xy}-1)+u$ ($1\leq u\leq11$).
The total number $N$ of nodes for one team is 198 ($=18\times11$). 

\begin{figure}[H]
	\centering
		\includegraphics[width=7cm]{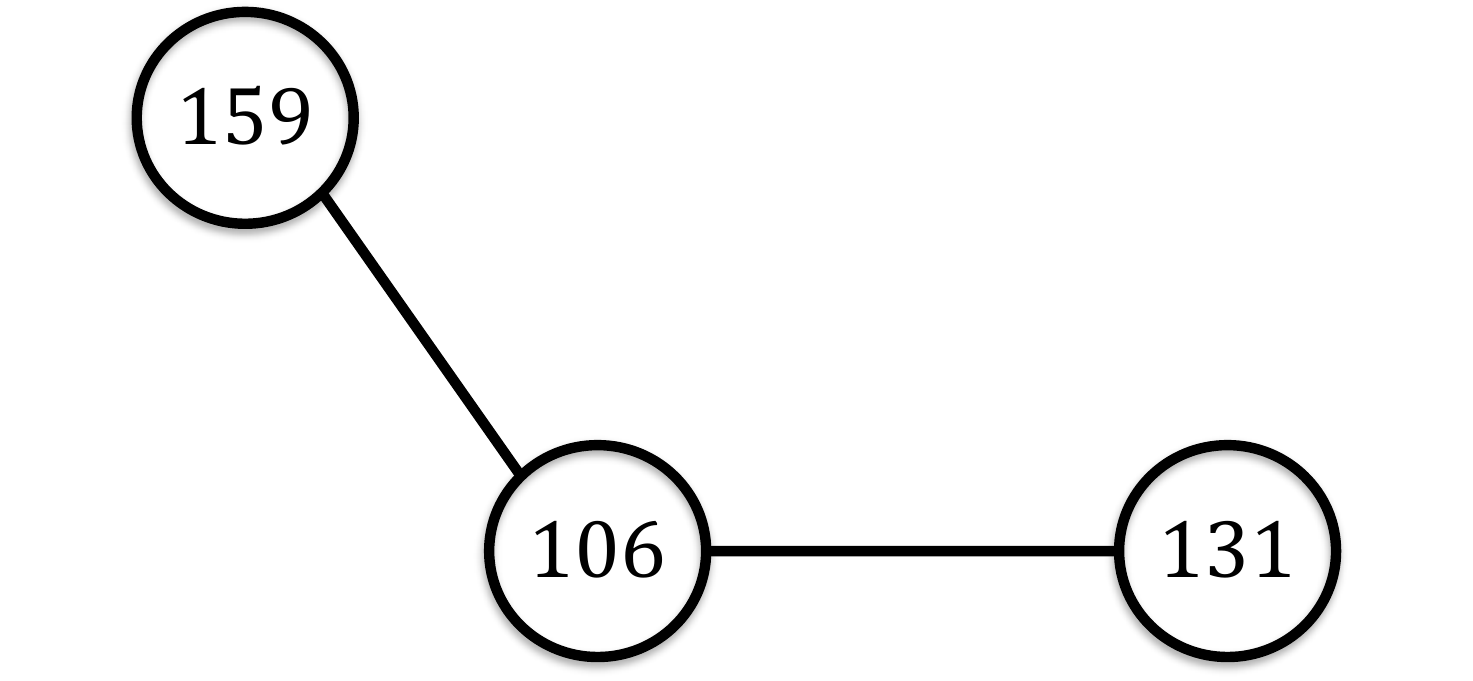}
		\caption{The network obtained from Fig. \ref{fig:field}. Each node corresponds to the three players in Fig. \ref{fig:field}: ``$159(u=5)$", ``$106(u=7)$", and ``$131(u=10)$".}
		\label{fig:n_example}
\end{figure}

When one player passes the ball to another player in the game, two nodes corresponding to these two players are connected by an undirected edge.
If more than one passes are made between the same nodes, multiple edges are allowed.
A ball-passing network is obtained as the set of the all passes made by one team in a game.
To be precise, we use the following rules.
 (i) Only the passes between players belonging to the same team are considered.
 (ii) When a player is replaced by a reserved player, the node for the new player is given the same number as the old player.
From Fig. \ref{fig:field}, for example,
we obtain the network among the three nodes ``$159(u=5)$", ``$106(u=7)$", and ``$131(u=10)$" as shown in Fig. \ref{fig:n_example}.

We focus on the following basic properties characterizing the network structure: 
the total number $M$ of edges,
the degree distribution $f(k)$,
the average degree $\langle k\rangle$,
mean path length $\ell$,
and clustering coefficient $C$.
Here, we consider a network with multiple edges in calculation of $k$ and $\langle k \rangle$, and a network with single edges in calculation of $\ell$ and $C$.

\section{Analysis of real data}
\subsection{Structural properties of the networks}
\begin{figure}[H]
	\begin{minipage}{.5\textwidth}
		\centering
		\includegraphics[width=8cm]{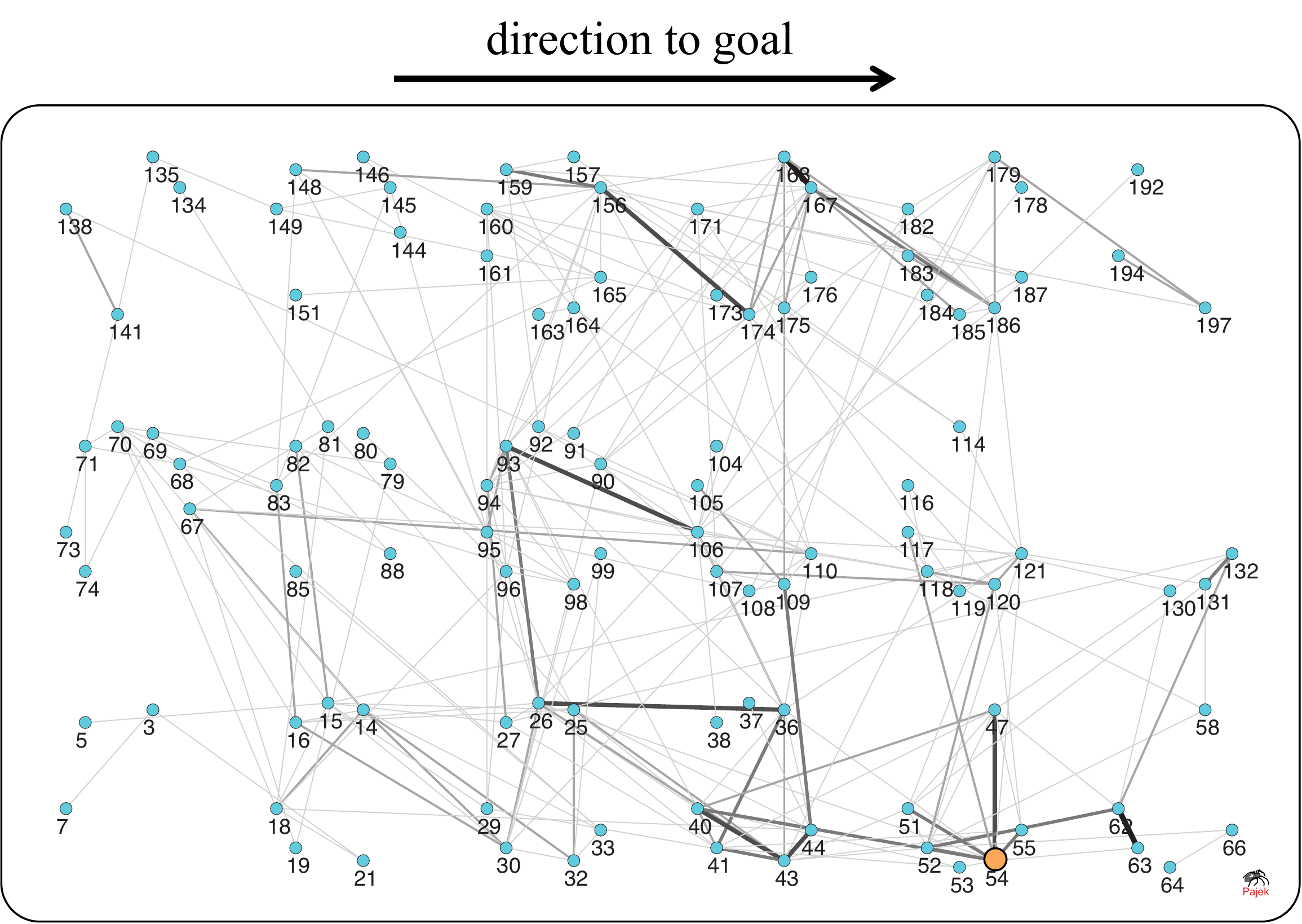}
		\caption*{(a) Japan}
	\end{minipage}%
		\begin{minipage}{.5\textwidth}
			\centering
			\includegraphics[width=8cm]{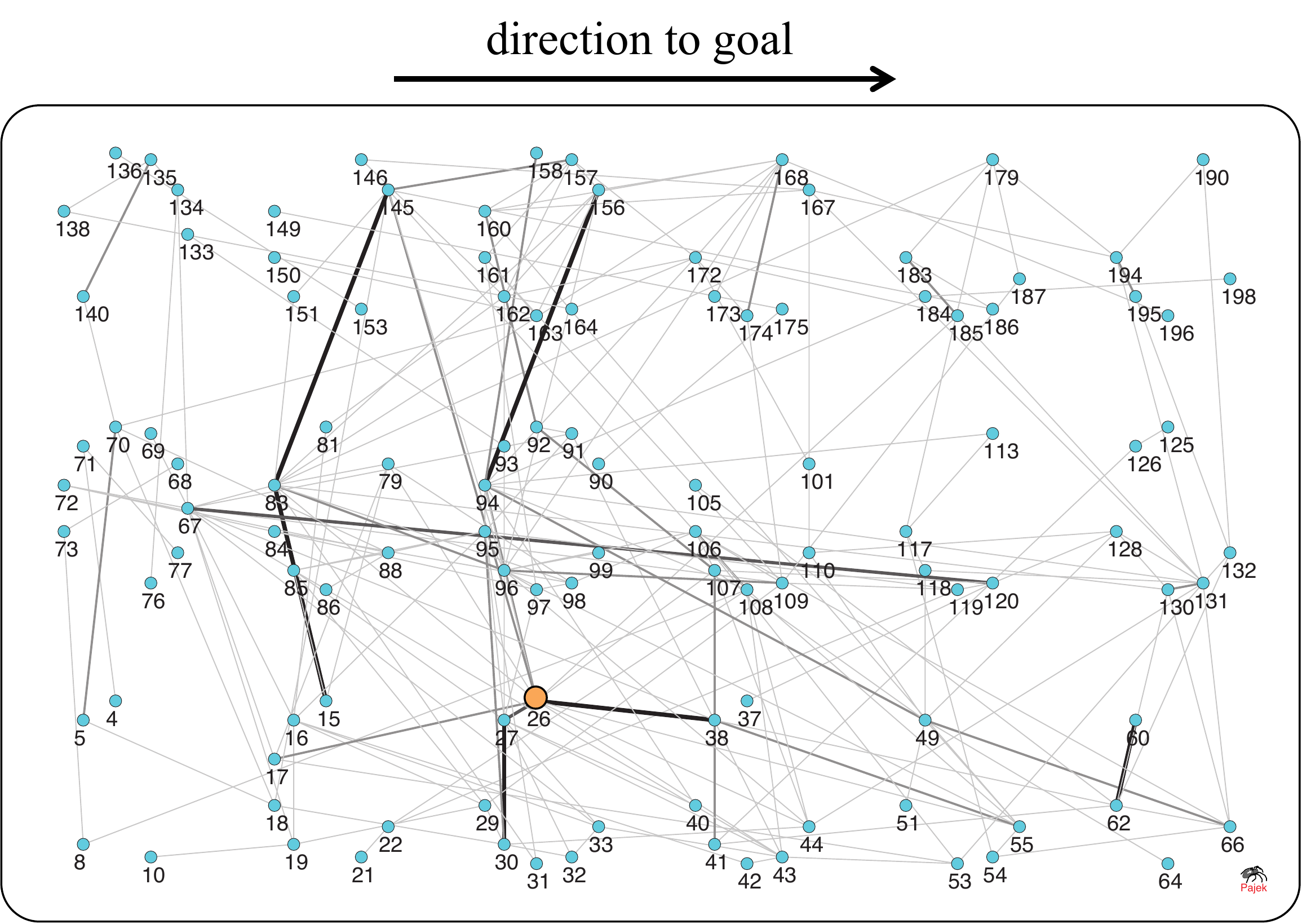}
			\caption*{(b) North Korea}
	\end{minipage}
	\caption{Network diagrams obtained from the game (iii). The darker edges represent higher frequency of passing. The yellow circle of each panel represents the hub node which has the maximum degree in the network. (These figures are created by means of \textit{Pajek} \cite{pajek}.) }
	\label{fig:net}
\end{figure}

We analyzed the five football games summarized in Table \ref{tb:game_data}.
The network diagrams of Japan and North Korea in the game (iii) are shown in Fig. \ref{fig:net}.
In these networks, all nodes are set on their own areas (the isolated nodes are excluded).
The edges are drawn in grayscale depending on the multiplicity; 
the darker one represents higher frequency of passing.
The direction of offense is rightward in both panels.
We find some hubs (node ``54" in Fig. \ref{fig:net} (a) and node ``26" in Fig. \ref{fig:net} (b) for example).
\begin{table}[H]
	\centering
 	\caption{Real game data for the analysis.}
	\vspace*{-0.3cm} 
	\label{tb:game_data}
 		\begin{tabular}{cccccc}
		\toprule
		     & Game       & Place & Date       & Score &  Competition             \\ \toprule
		(i)  & Japan vs Vietnam       & Japan           & 2011.10.07 & 1-0   & Kirin Challenge Cup    \\
		(ii) & Japan vs Tajikistan      & Japan           & 2011.10.11 & 8-0   & WCup Asian qualifier  \\
		(iii) & Japan vs North Korea & North Korea   & 2011.11.15 & 0-1   & WCup Asian qualifier  \\ 
		(iv) & Spain vs Italy            & Poland           & 2012.06.10 & 1-1   & Euro 2012                \\ 
		(v) & Germany vs Holland    & Ukraine         & 2012.06.13 & 2-1   & Euro 2012                 \\
		\bottomrule 
		\end{tabular}
\end{table}

Characteristic values of each network are summarized in Table \ref{tb:character-real}.
The mean path length $\ell$ and clustering coefficient $C$ are compared with $\ell_{\rm rand}$ and $C_{\rm rand}$ of random networks having the same $N$ and $M$; they were calculated by averaging over 1000 samples.
It is found that the sum of the edges of the two teams is almost constant in the five games, 
although $M$ of each team largely varied.
Since the time of a football game is 90 minutes in typical,
the time elapsing from one node receive the ball and the node pass it to another
seemed to be almost constant for all games.
The mean path length $\ell$ is about $3.3\pm 0.3$,
and is close to $\ell_{\rm rand}=4.4\pm 1.7$.
The clustering coefficient $C$ of each network is about ten times greater than that of random network $C_{\rm rand}$. 
Therefore, these ball-passing networks are considered to possess small-world property.

\begin{table}[H]
  \centering
  \caption{Characteristic values of each network.}
  \vspace*{-0.3cm} 
    \begin{tabular}{cccccccc}
    \toprule
    Game                          & Team            &  $M$ & $\langle k \rangle$ & $\ell$  & $\ell_{\rm rand}$ & $C$  & $C_{\rm rand}$ \\ \toprule
    \multirow{2}[2]{*}{(i)}    & Japan           &  546 & 5.5   & 3.07  & 3.29 & 0.25  & 0.027 \\
                                       & Vietnam       &  278 & 2.8    & 3.42  & 5.00 & 0.21   & 0.014 \\ \midrule
    \multirow{2}[2]{*}{(ii)}    & Japan          & 751 & 7.6 & 2.75  & 2.84 & 0.36  & 0.038 \\
                                       & Tajikistan     &  139 & 1.4    & 3.66 & 8.53 & 0.11  &  0.007\\ \midrule
    \multirow{2}[2]{*}{(iii)}   & Japan           &  353 & 3.6  & 3.52  &  4.23 & 0.18  &  0.018\\
                                       & North Korea & 272 & 2.7    & 3.49  & 5.08 & 0.11  &  0.014\\ \midrule
    \multirow{2}[2]{*}{(iv)}   & Spain           &  688 & 6.9 & 3.11  &  2.95 & 0.29  &  0.035\\
                                       & Italy              &  320 & 3.2  & 3.43  & 4.52 & 0.17  &  0.016\\ \midrule
    \multirow{2}[2]{*}{(v)}    & Germany      &  428 & 4.3  & 3.39  &   3.76 & 0.18  &  0.022\\
                                        & Holland        &  501 & 5.1  & 3.05  &  3.44 & 0.23  &  0.025\\ 
    \bottomrule
    \end{tabular}
    \label{tb:character-real}
\end{table}

\subsection{Degree distribution and Edge-multiplicity distribution}
\begin{figure}[H]
	\begin{minipage}{.5\textwidth}
		\centering
		\includegraphics[width=8cm]{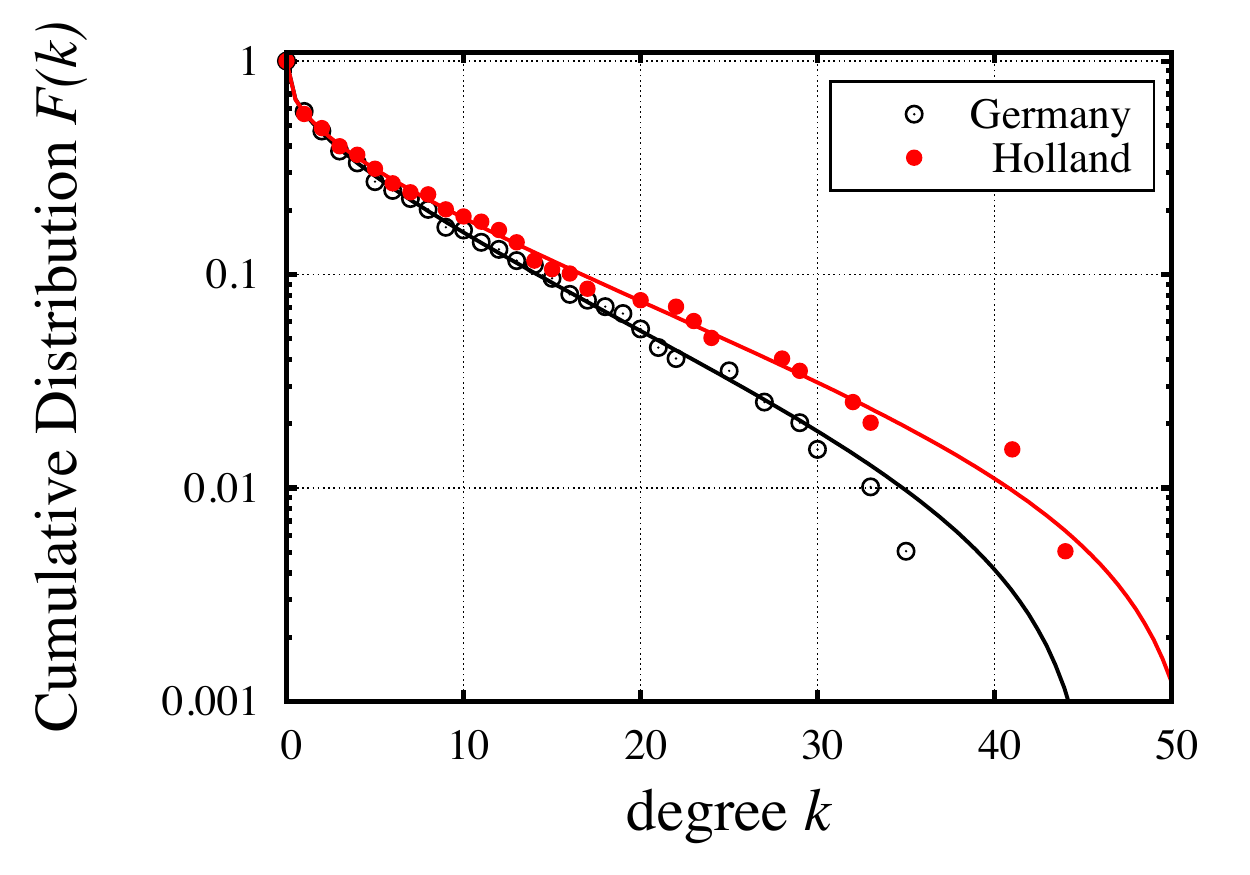}
		\caption*{(v)}
	\end{minipage}
	\caption{The cumulative degree distribution of each network. The panel (i)-(v) correspond to the five games in Table \ref{tb:game_data}. Each distribution is in good agreement with the truncated gamma distribution given as Eq. \eqref{eq:cdf}.}
	\label{fig:degree}
\end{figure}

The cumulative degree distributions of the networks are shown in Fig. \ref{fig:degree}
with single logarithmic scale.
The solid curves in each panel are truncated gamma distributions.
Here, the probability density function $f(k)$ and cumulative distribution function $F(k)$ 
of the truncated gamma distribution are given as follows:
\begin{eqnarray}
\label{eq:pdf}
f(k) &=& \frac{1}{\lambda^\nu\gamma (\nu,k_{\rm max}/\lambda,0)}k^{\nu-1}e^{-\frac{k}{\lambda}}, \\
\nonumber \\
\label{eq:cdf}
F(k) &=& \frac{\gamma (\nu,k_{\rm max}/\lambda,k/\lambda)}{\gamma(\nu,k_{\rm max}/\lambda,0)} ,
\end{eqnarray}
where $\gamma(\nu,k_{\rm max}/\lambda,k/\lambda)$ denotes the upper incomplete gamma function with the upper limit of integration $k_{\rm max}/\lambda$, defined as
\begin{equation}
\label{eq:in-gamma}
	\gamma(\nu,k_{\rm max}/\lambda,k/\lambda) = \int^{k_{\rm max}/\lambda}_{k/\lambda}t^{\nu-1}e^{-t}dt .
\end{equation}
The domain of $k$ is $0\le k \le k_{\rm max}$.
$\nu$, $\lambda$, and $k_{\rm max}$ are fitting parameters for the real data. 
From Eq. \eqref{eq:pdf}, $\nu$ controls the shape of $f(k)$ exhibiting the power-law behavior, and $\lambda$ controls the crossover point between the power-law and the exponential behaviors.
The truncated gamma distribution has been used for a life test \cite{Johnson} and precipitation amount \cite{Das1955} for example.

As shown in Fig. \ref{fig:degree}, $F(k)$ for each team can be fitted quite well by selecting the values of $\nu$, $\lambda$, and $k_{\rm max}$ in Eqs. \eqref{eq:cdf} and \eqref{eq:in-gamma}.
Actually, the values of $\nu$, $\lambda$ and $k_{\rm max}$ for fittings are shown in Table \ref{tab:fitting-real}.
It is found that the values of $\nu$ of each network is about $0.34\pm 0.07$, and the curve exhibits power-law behavior in the low degree part, and exponential behavior in high degree part (see Eq. \eqref{eq:pdf} for reference).

\begin{figure}[H]
	\begin{minipage}{.5\textwidth}
		\centering
		\includegraphics[width=8cm]{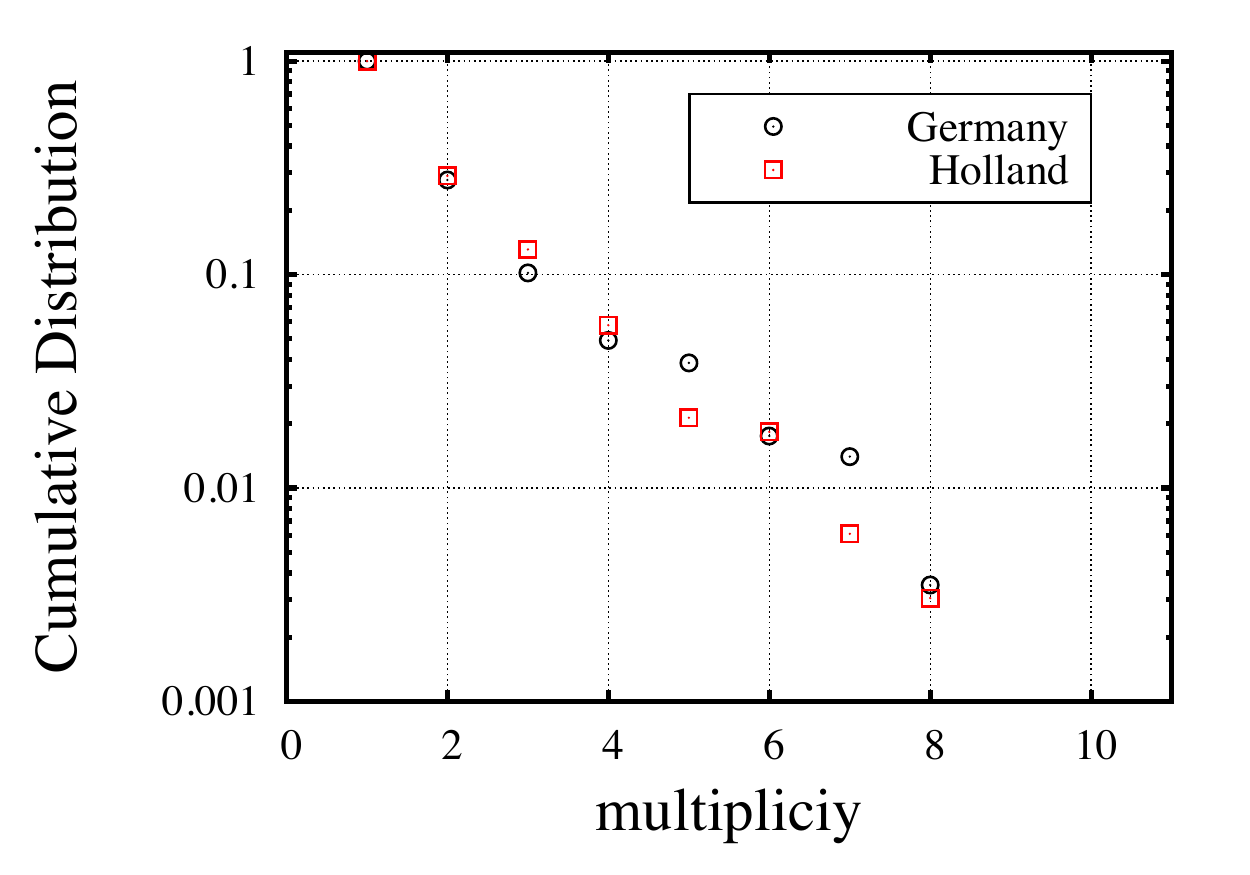}
		\caption*{(v)}
	\end{minipage}
	\caption{The cumulative edge-multiplicity distribution of each network. The panels (i)-(v) correspond to the five games in Table \ref{tb:game_data}.}
	\label{fig:multiple}
\end{figure}

The cumulative edge-multiplicity distributions of the networks are shown in Fig. \ref{fig:multiple} with single logarithmic scale. 
It is found that each distribution decreases almost exponentially.
Then almost $70\%$ of edges have multiplicity one i.e., single edges, and there are a few high multiple edges.

\clearpage

\begin{figure}[H]
	\begin{minipage}{.5\textwidth}
		\centering
		\includegraphics[width=8cm]{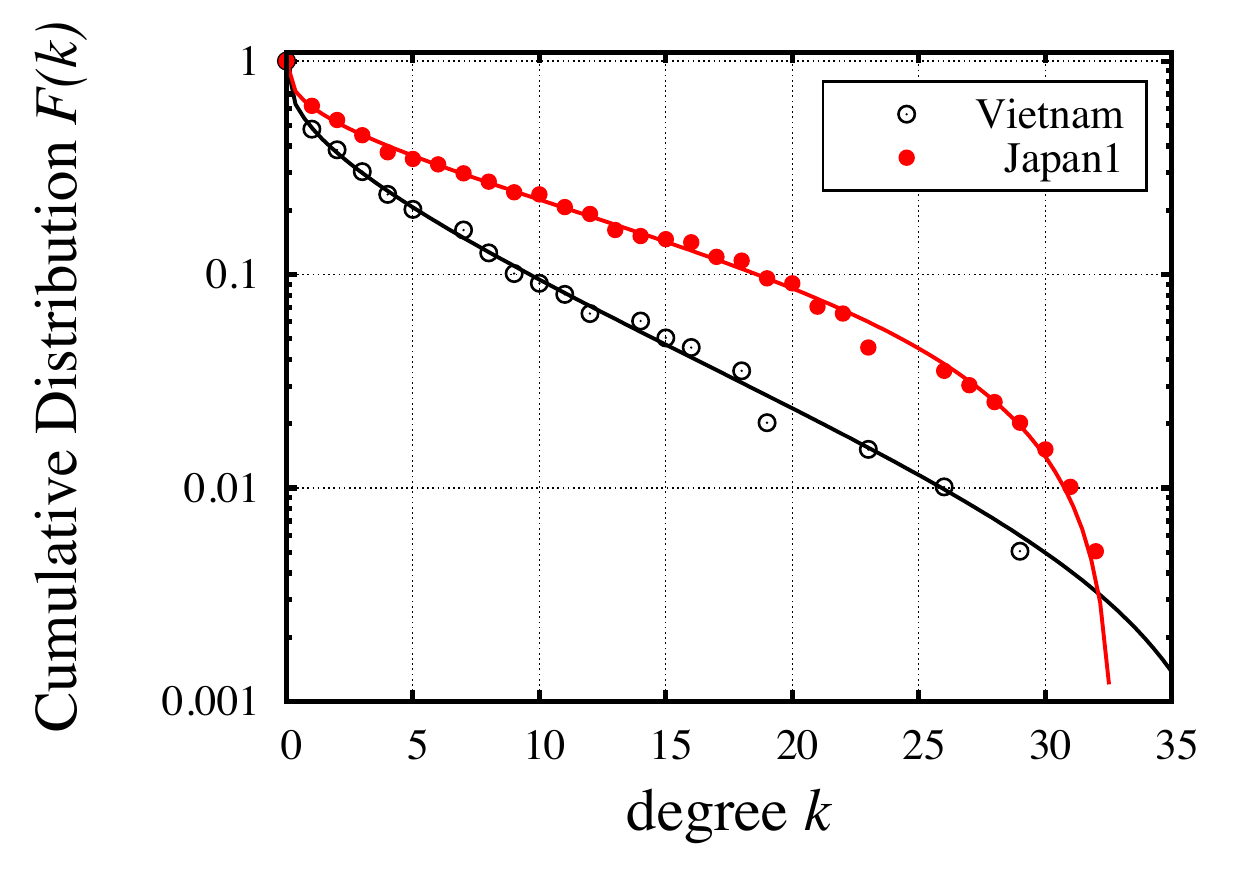}
		\caption*{(i)}
	\end{minipage}%
	\begin{minipage}{.5\textwidth}
		\centering
		\includegraphics[width=8cm]{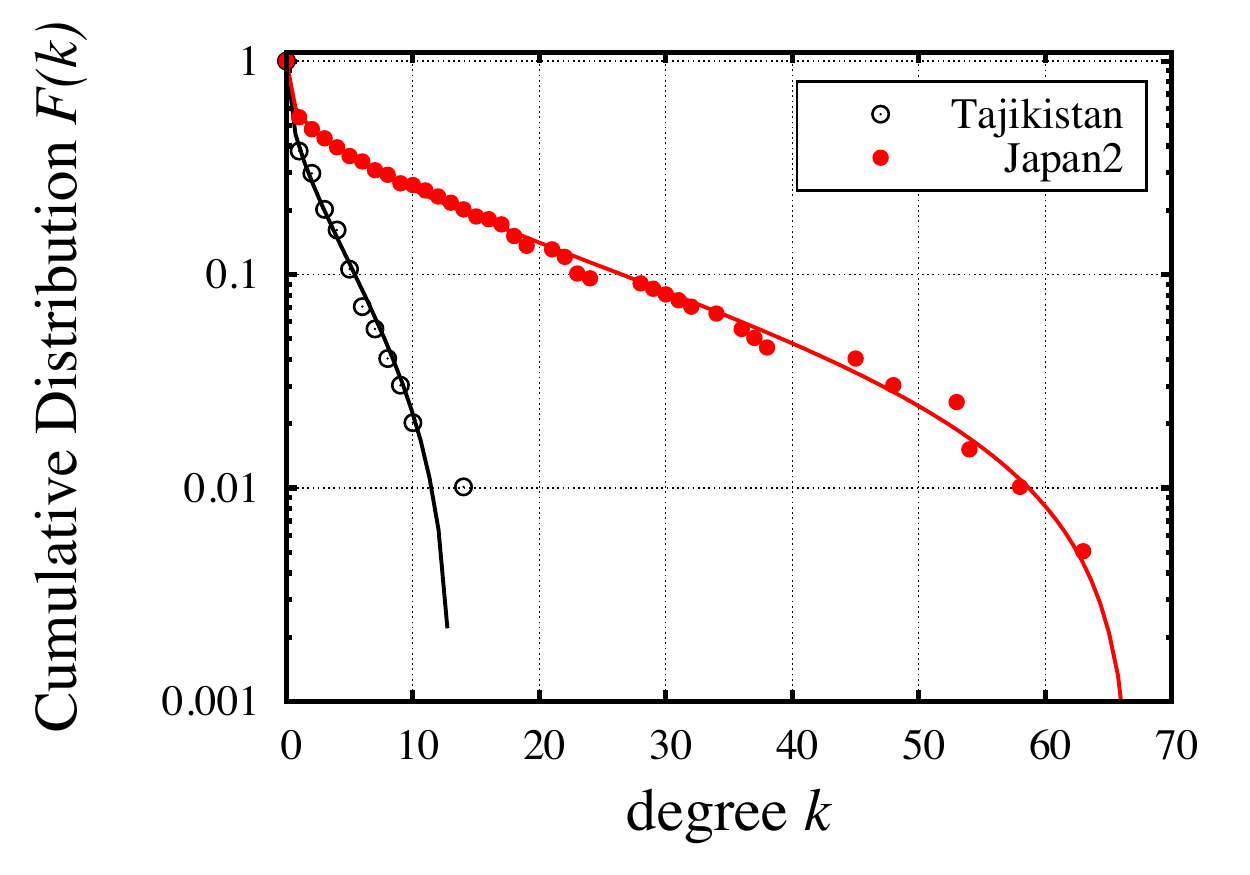}
		\caption*{(ii)}
	\end{minipage}
\end{figure}

\begin{figure}[H]
	\begin{minipage}{.5\textwidth}
		\centering
		\includegraphics[width=8cm]{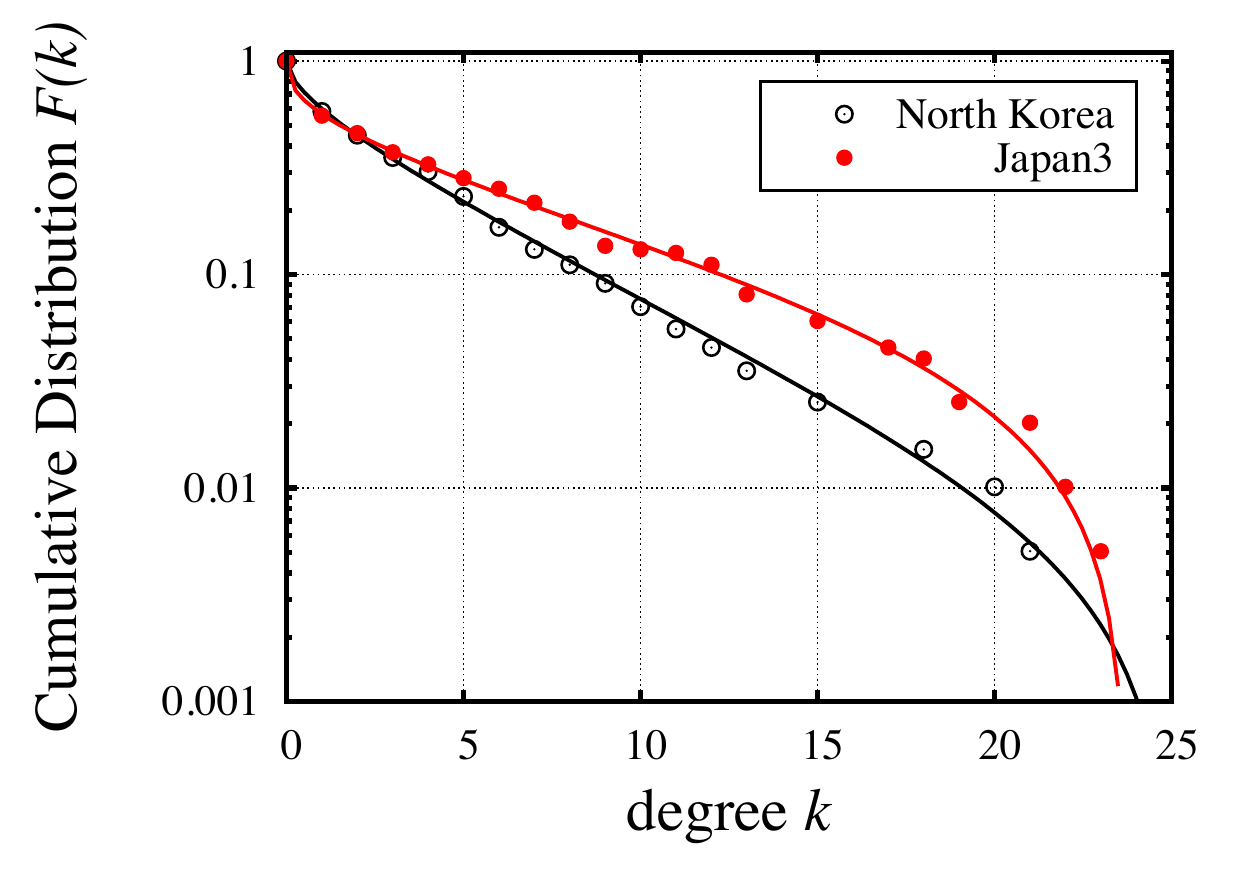}
		\caption*{(iii)}
	\end{minipage}%
	\begin{minipage}{.5\textwidth}
		\centering
		\includegraphics[width=8cm]{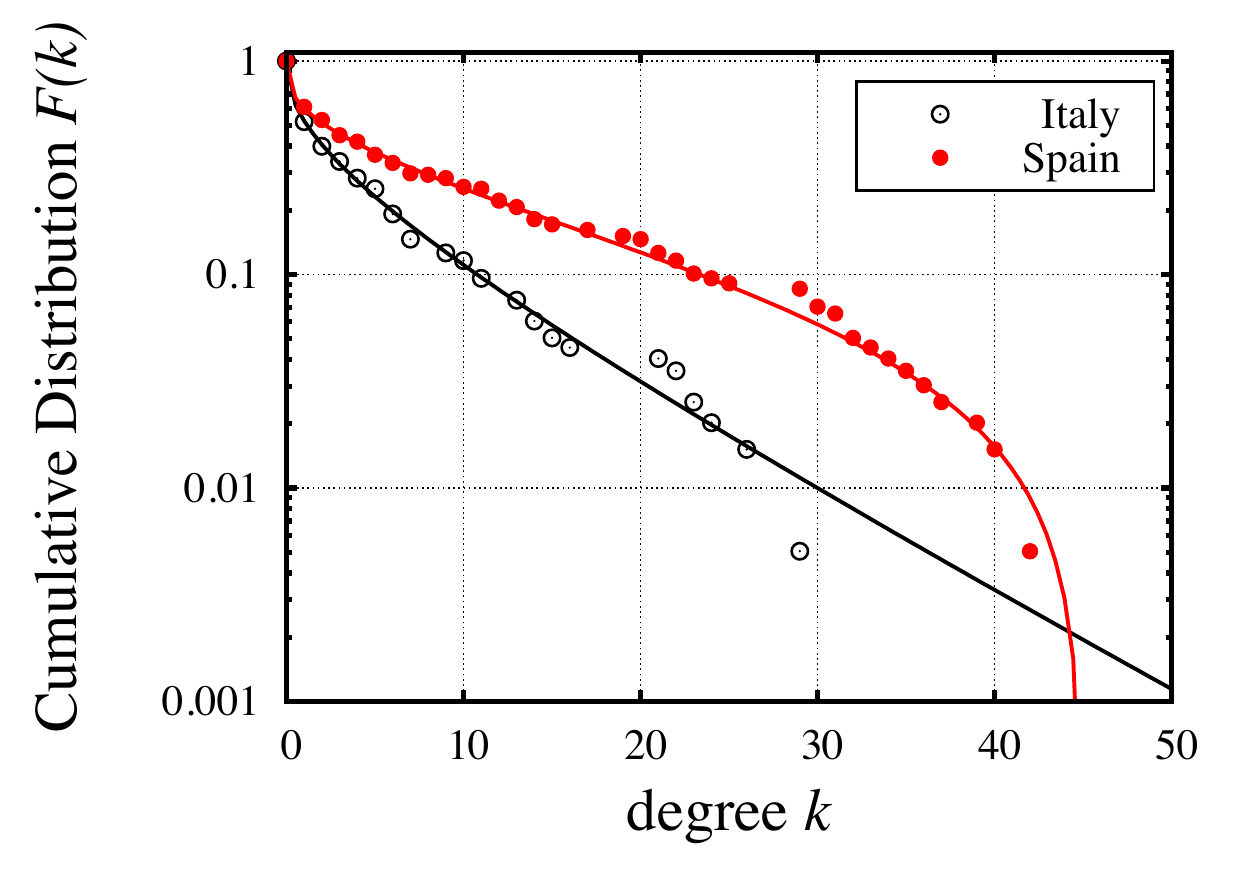}
		\caption*{(iv)}
	\end{minipage}
\end{figure}

\begin{table}[H]
  \centering
  \caption{The parameters for fittings of each network.}
  \vspace*{-0.3cm} 
    \begin{tabular}{cccccc}
    \toprule
              Game                   & Team              & $\nu$    & $\lambda$    & $k_{\rm max}$ \\ \toprule
    \multirow{2}[2]{*}{(i)}    & Japan              & 0.33 & 34.6   & 32.8     \\ 
                                          & Vietnam          & 0.32 & 10.4   & 38.1      \\ \midrule
    \multirow{2}[2]{*}{(ii)}    & Japan              & 0.26 & 43.3   & 67.0          \\
                                          & Tajikistan        & 0.31 & 6.9   & 13.1    \\ \midrule
    \multirow{2}[2]{*}{(iii)}   & Japan              & 0.35 & 16.9   & 23.7       \\
                                           & North Korea   & 0.53 & 6.3   & 25.0            \\ \midrule
    \multirow{2}[2]{*}{(iv)}    & Spain             & 0.29 & 44.0 & 45.0       \\
                                           & Italy                & 0.35 & 10.6   & 185.7          \\ \midrule
    \multirow{2}[2]{*}{(v)}     & Germany        & 0.36 & 14.3  & 45.9        \\
                                            & Holland          & 0.33 & 18.3  & 51.9         \\
    \bottomrule
    \end{tabular}%
  \label{tab:fitting-real}%
\end{table}%

\clearpage

\begin{figure}[H]
	\begin{minipage}{.5\textwidth}
		\centering
		\includegraphics[width=8cm]{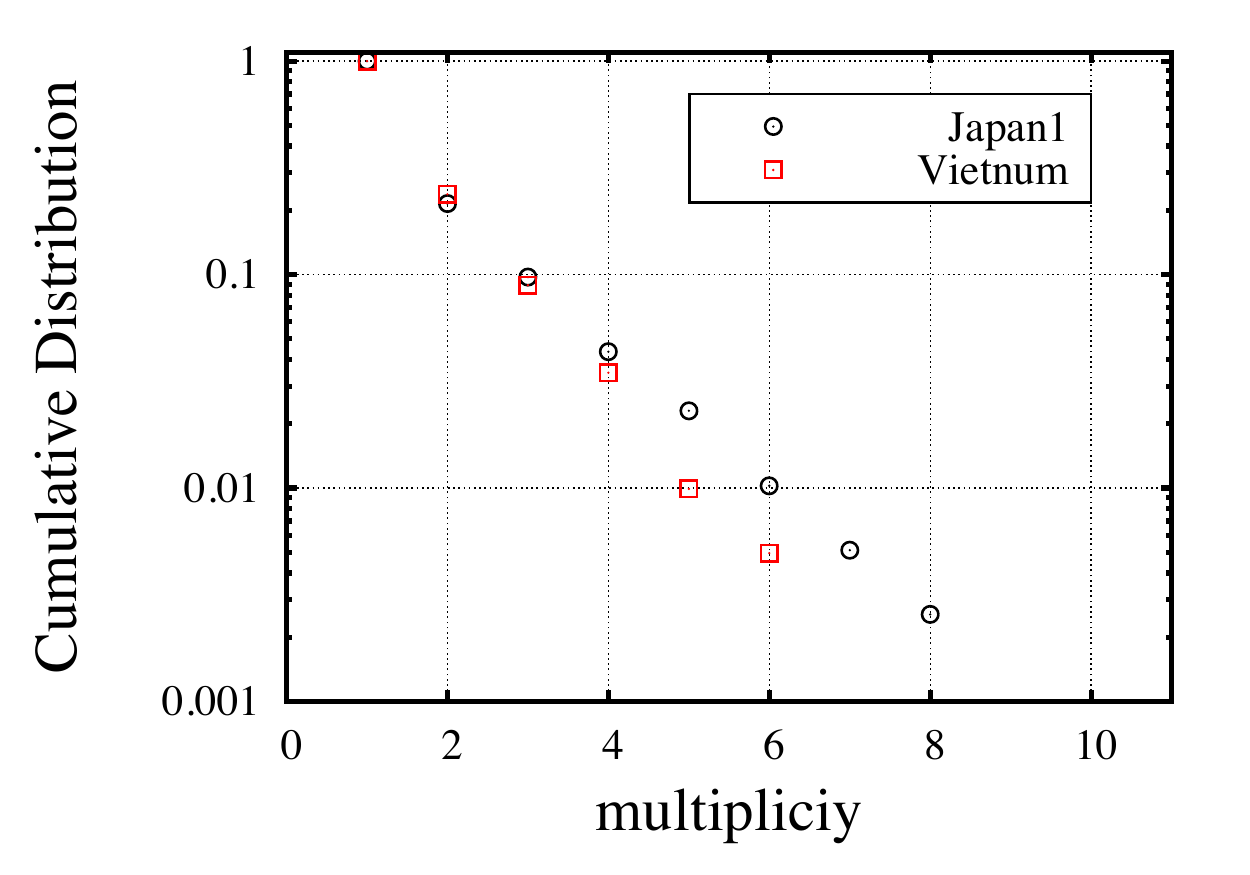}
		\caption*{(i)}
	\end{minipage}
	\begin{minipage}{.5\textwidth}
		\centering
		\includegraphics[width=8cm]{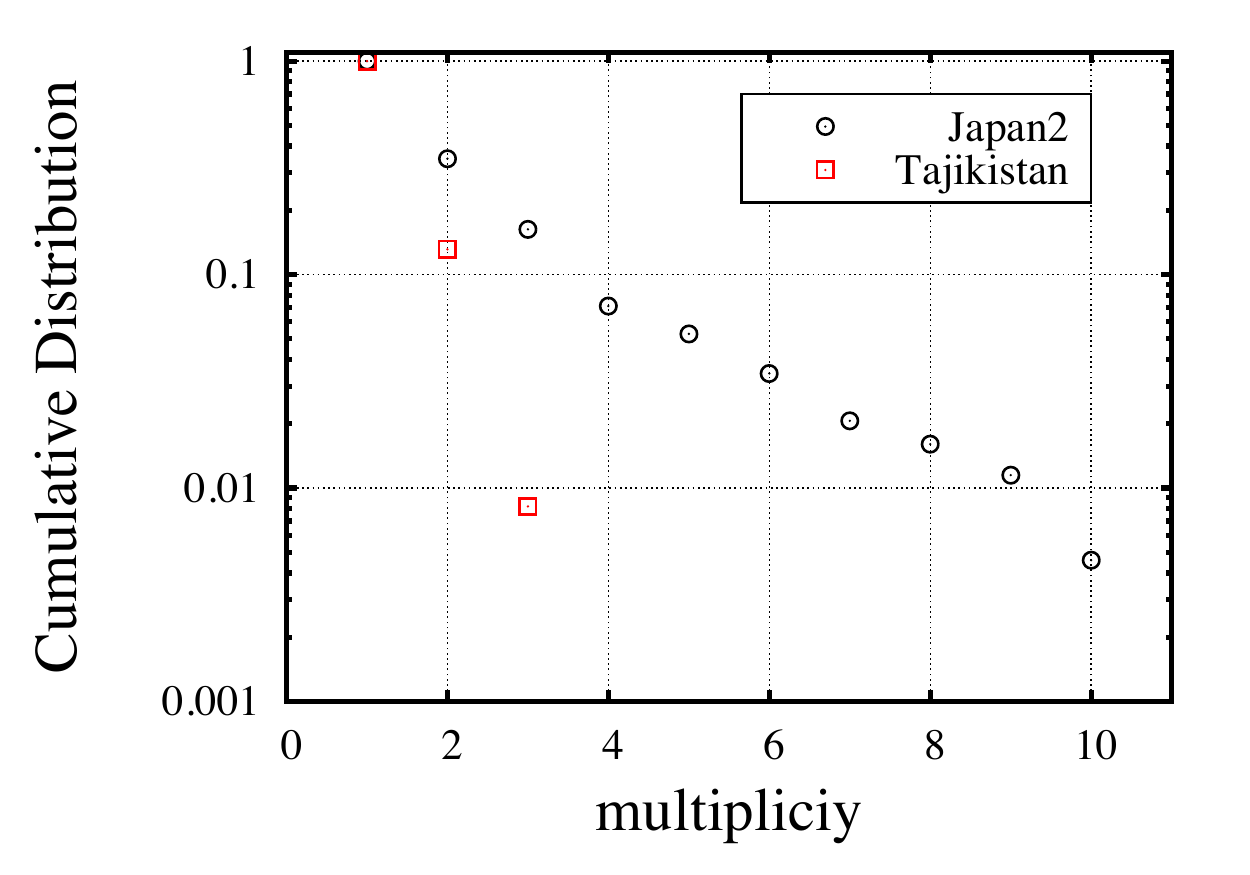}
		\caption*{(ii)}
	\end{minipage}
\end{figure}

\begin{figure}[H]
	\begin{minipage}{.5\textwidth}
		\centering
		\includegraphics[width=8cm]{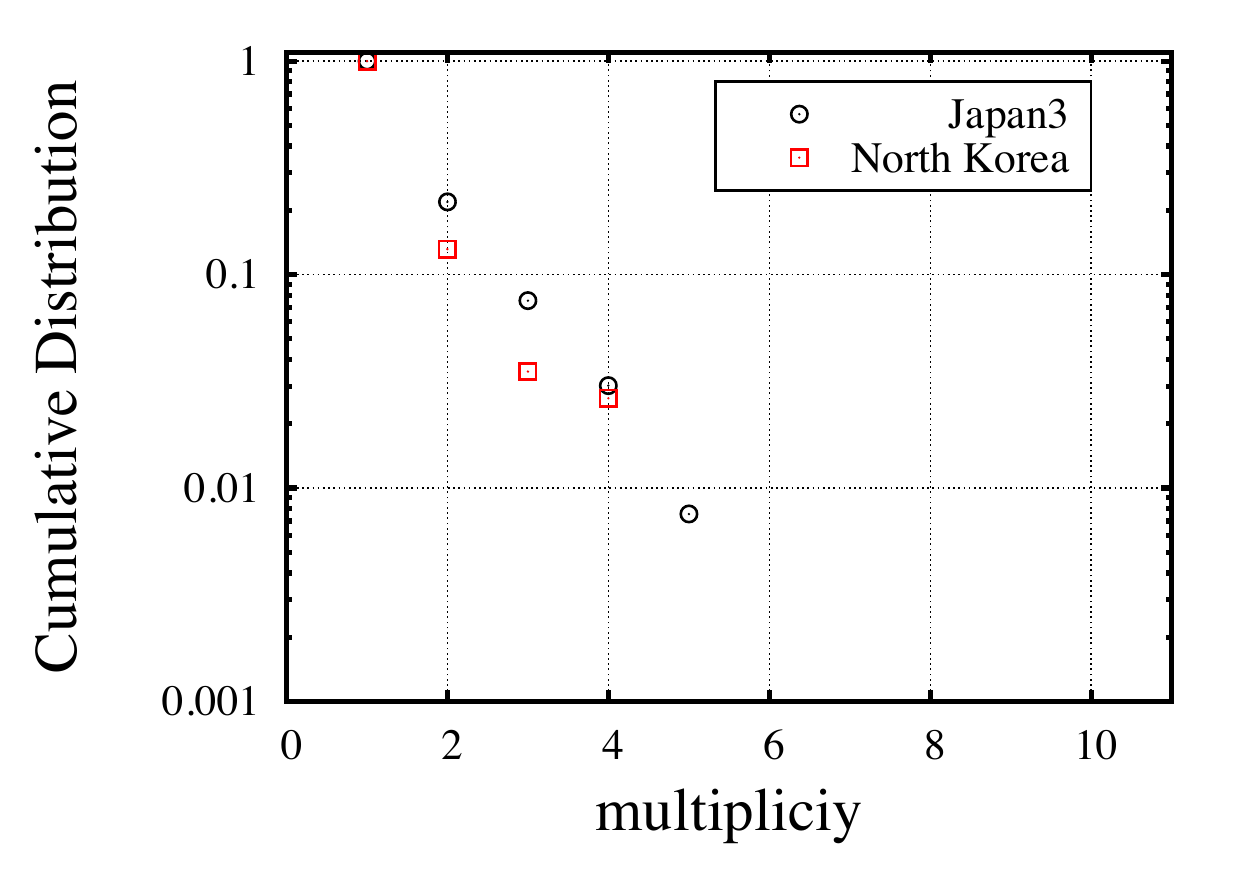}
		\caption*{(iii)}
	\end{minipage}
	\begin{minipage}{.5\textwidth}
		\centering
		\includegraphics[width=8cm]{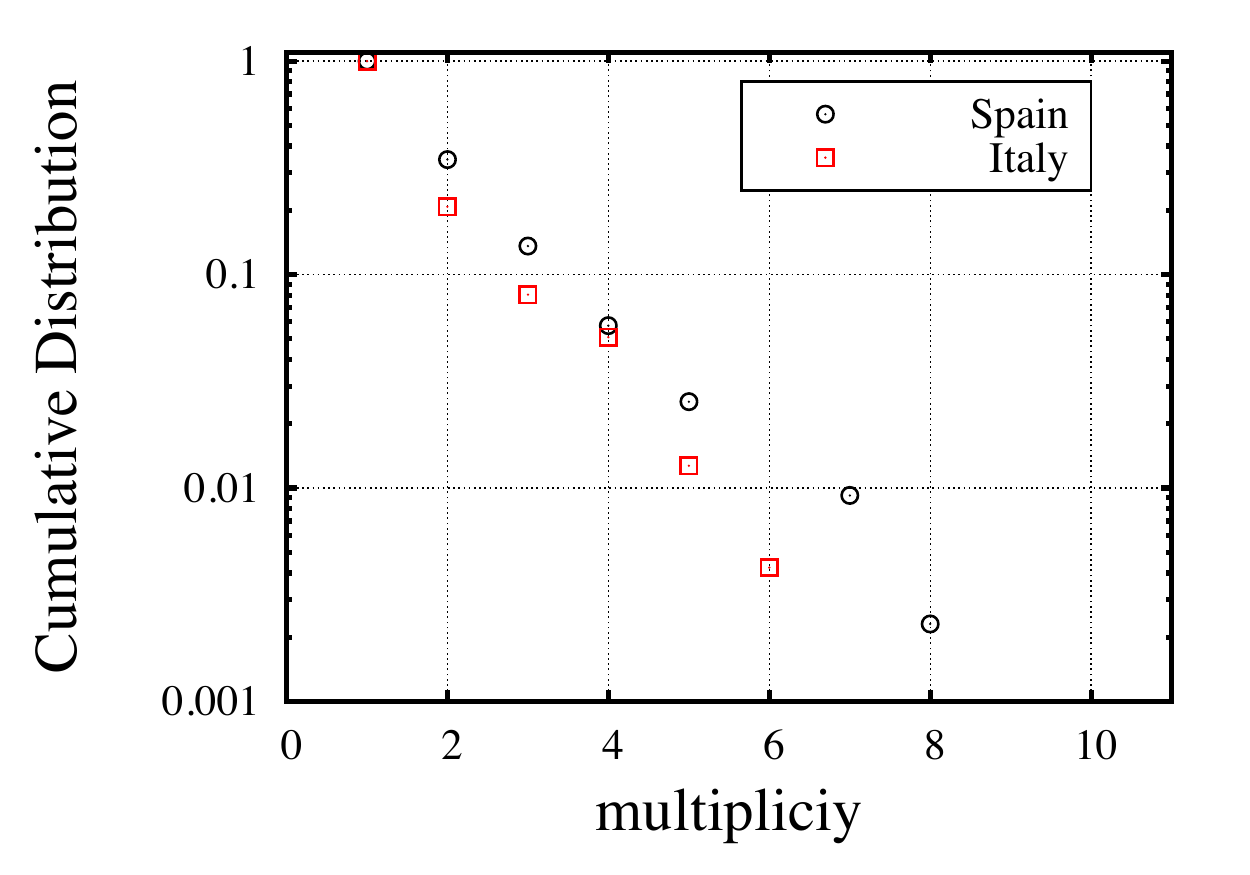}
		\caption*{(iv)}
	\end{minipage}
\end{figure}

\clearpage

\section{Modeling and numerical calculation}
In the ball-passing network, the degree of each node represents the sum of the numbers of passing and receiving.
And the sum corresponds to the frequency of ball possession.
Therefore, the degree distribution shown in section 3 can be interpreted as the distribution of ball-possession frequency.
In this section, we assume the sequence of ball passing as a random motion of a ball between nodes for simplicity.
And, we try to reproduce the degree distribution by focusing on ``ball-possession probability" with a Markov-chain model.

\subsection{Markov-chain model}
We define $N$ nodes in the same way as the section 2,
where we treat the two types of field division in order to
take influence of difference in field division on the statistical properties into account.
Actually, we show the cases of $N=198$ and 3168 for $3\times6$ and $12\times24$ divisions, respectively.

Now, we introduce
the ball-possession probability $a_{i}^{(t)}$ for the node $i$ at time $t$,
and the probability vector $\bm{a}^{(t)}$ is defined as
\begin{equation}
	\bm{a}^{(t)}=[a_{1}^{(t)},a_{2}^{(t)},\ldots,a_{i}^{(t)},\ldots,a_{N}^{(t)}].
\end{equation}
Here, this probability vector is normalized as
\begin{equation}
	\sum_{i=1}^{N}a_{i}^{(t)}=1.
\end{equation}
We assume that the time evolution of $\bm{a}^{(t)}$ is given as the Markov chain 
\begin{equation}
	\bm{a}^{(t+1)}=\bm{a}^{(t)}\bm{P},
\end{equation}
where $\bm{P}$ is a transition matrix with $N \times N$ elements.
Each element $P_{i\to j}$ of $\bm{P}$ gives
the ball-passing probability from the node $i$
to the node $j$.
The transition matrix is also normalized as
\begin{equation}
\label{eq:normalize}
	\sum_{j=1}^{N}P_{i\to j}=1
\end{equation}
for each $i$.
Note that the transition matrix is not necessarily symmetric.

A single time step of the Markov chain is a stochastic expression of a single passing event.
And $a_{i}^{(t)}$ is the average ball-possession probability of node $i$ after $t$ passing events.
(Note that this ``average" is the ensemble average for many samples).
We assume that the ball-passing statistics after $t$ passing events in the real data
correspond to the $\bm{a}^{(t)}$ with the same $t$.

For the initial condition of the time evolution, which corresponds to the kick-off in football games,
we consider the case where the ball passing begins with the player 11
at area $(3,2)$ in $3\times6$, and at $(12,6)$ in $12\times24$.
This condition corresponds to the case that
the 98th (1452nd) element of $\bm{a}^{(0)}$ has probability 1 for $3\times6$ ($12\times24$)
and the others have probability 0.

\begin{figure}[H]
	\begin{minipage}{.5\textwidth}
		\centering
		\includegraphics[width=7cm]{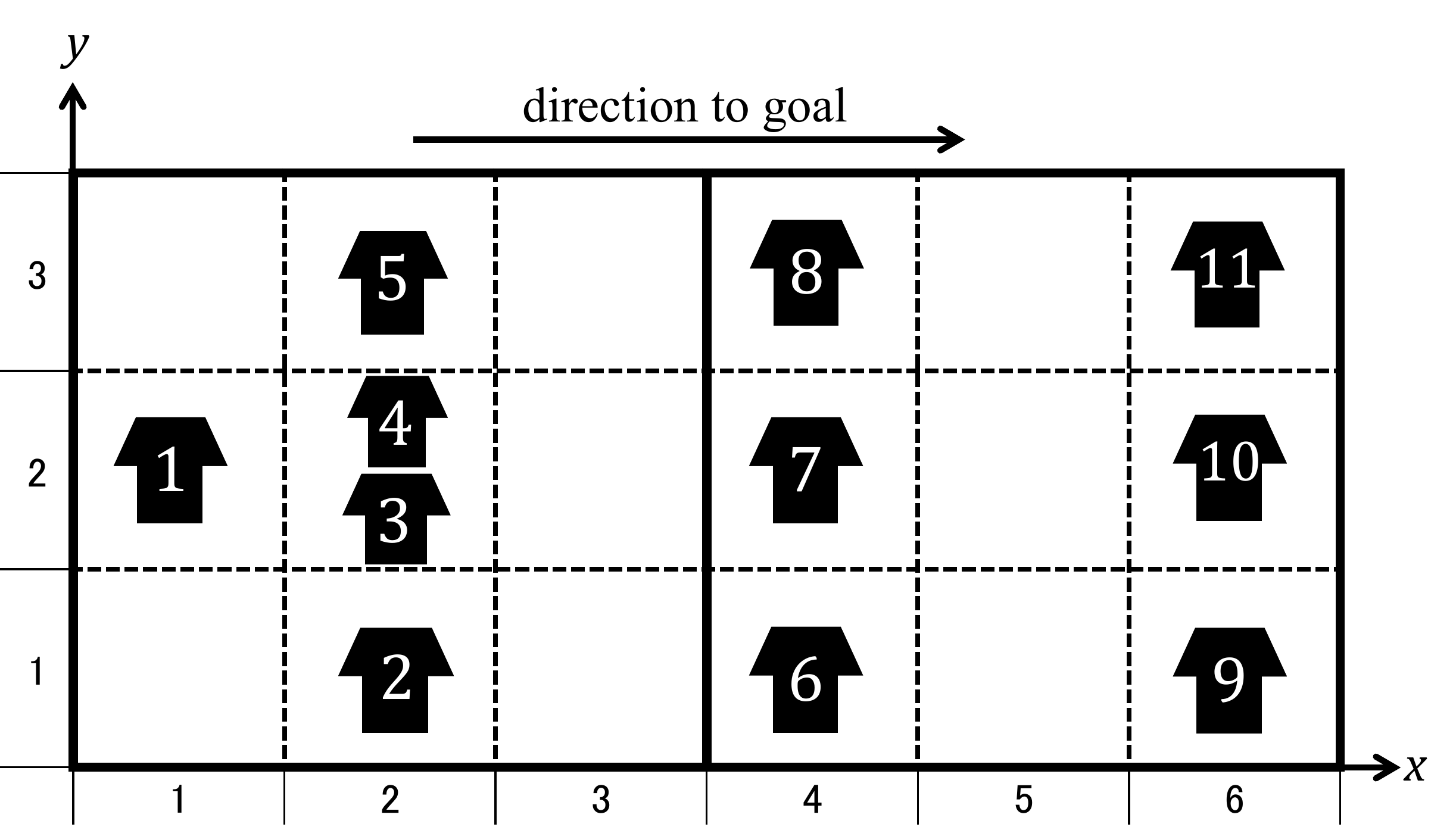}
		\caption*{(i)}
	\end{minipage}
	\begin{minipage}{.5\textwidth}
		\centering
		\includegraphics[width=7.7cm]{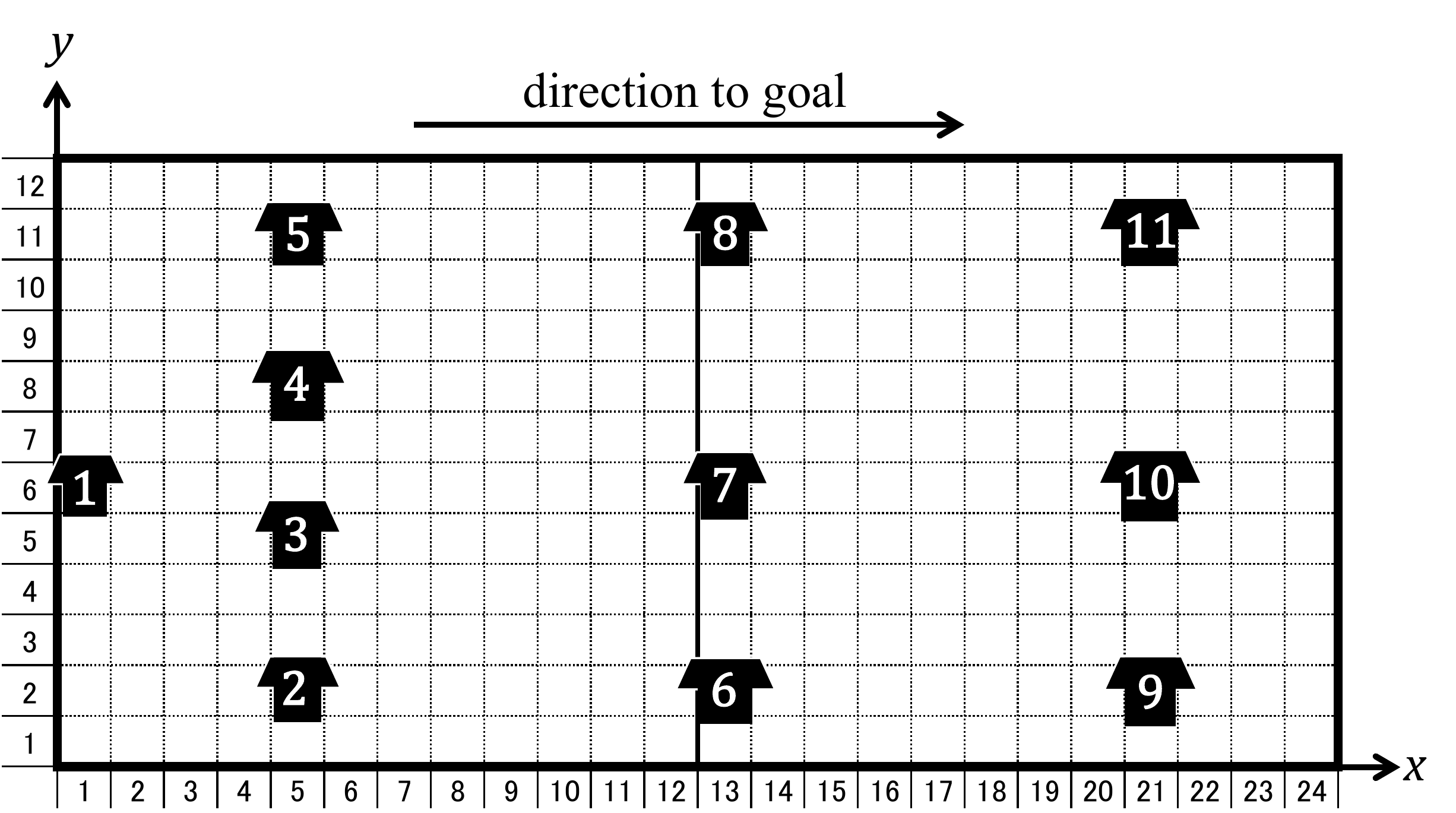}
		\caption*{(ii)}
	\end{minipage}
	\caption{The home position of each 11 player for the numerical simulation: (i) $3\times 6$, (ii) $12\times24$ divisions.}
	\label{fig:position}
\end{figure}

For the element $P_{i\to j}$, we assume the following.
(1) If the two nodes $i$ and $j$ represent the same players, then $P_{i\to j} = 0$; the dribble is not considered here.
(2) Otherwise, namely for $i$ and $j$ showing different players, $P_{i\to j}$ consists of the two factors:
\begin{equation}
\label{eq:Pij}
	P_{i\to j}\propto Q_{\alpha}(r_{ij})\times R_{\beta,\xi}(L_{j}).
\end{equation}
Here, $Q_{\alpha}(r_{ij})$ denotes the factor for the distance of passes, and represents the difficulty in completing a pass dependent on $r_{ij}$, the distance between the two nodes $i$ and $j$.
$R_{\beta,\xi}(L_{j})$ denotes the factor for existence probability
of the player receiving a pass.
$L_{j}$ represents the distance of the node $j$ from its home position as shown in Fig. \ref{fig:position}.
Normalization factor in Eq. \eqref{eq:Pij} is determined to satisfy Eq. \eqref{eq:normalize}.
For simplicity, we consider that $Q_{\alpha}(r_{ij})$ is a step function with a threshold distance $\alpha$
\begin{equation}
	\label{eq:Q}
	Q_{\alpha}(r_{ij})=
	\begin{cases}
		1&(r_{ij}\le \alpha)\\
		0&(r_{ij}>\alpha),\\
	\end{cases}\\
\end{equation}
and $R_{\beta,\xi}(L_{j})$ is an exponential function with a parameter $\beta$ and a threshold distance $\xi$
\begin{equation}
	\label{eq:R}
	R_{\beta,\xi}(L_{j})=
	\begin{cases}
		1&(L_{j}\le \xi)\\
		e^{-\beta(L_{j}-\xi)}&(L_{j}>\xi).\\
	\end{cases}
\end{equation}

\clearpage

\subsection{Numerical results}
We focus on the cumulative distribution $G(a)$ of $\bm{a}^{(500)}$ defined as the total number of $i$ which satisfies $a_{i}^{(500)} \ge a$, and compare it with the truncated gamma distribution.
And, we show the dependence of $G(a)$ on $Q_{\alpha}(r_{ij})$ and $R_{\beta,\xi}(L_{j})$. 
The reason why we set $t=500$ is as follows.
We have assumed that the degree of the node $i$ in the ball-passing network after $t$ passing events is proportional to $a_{i}^{(t)}$ with the same $t$.
And the number of passes made in one game is about 500 (see the values of $M$ in Table \ref{tb:character-real} for reference). 

\begin{figure}[H]
	\begin{minipage}{.5\textwidth}
		\centering
		\includegraphics[width=7.5cm]{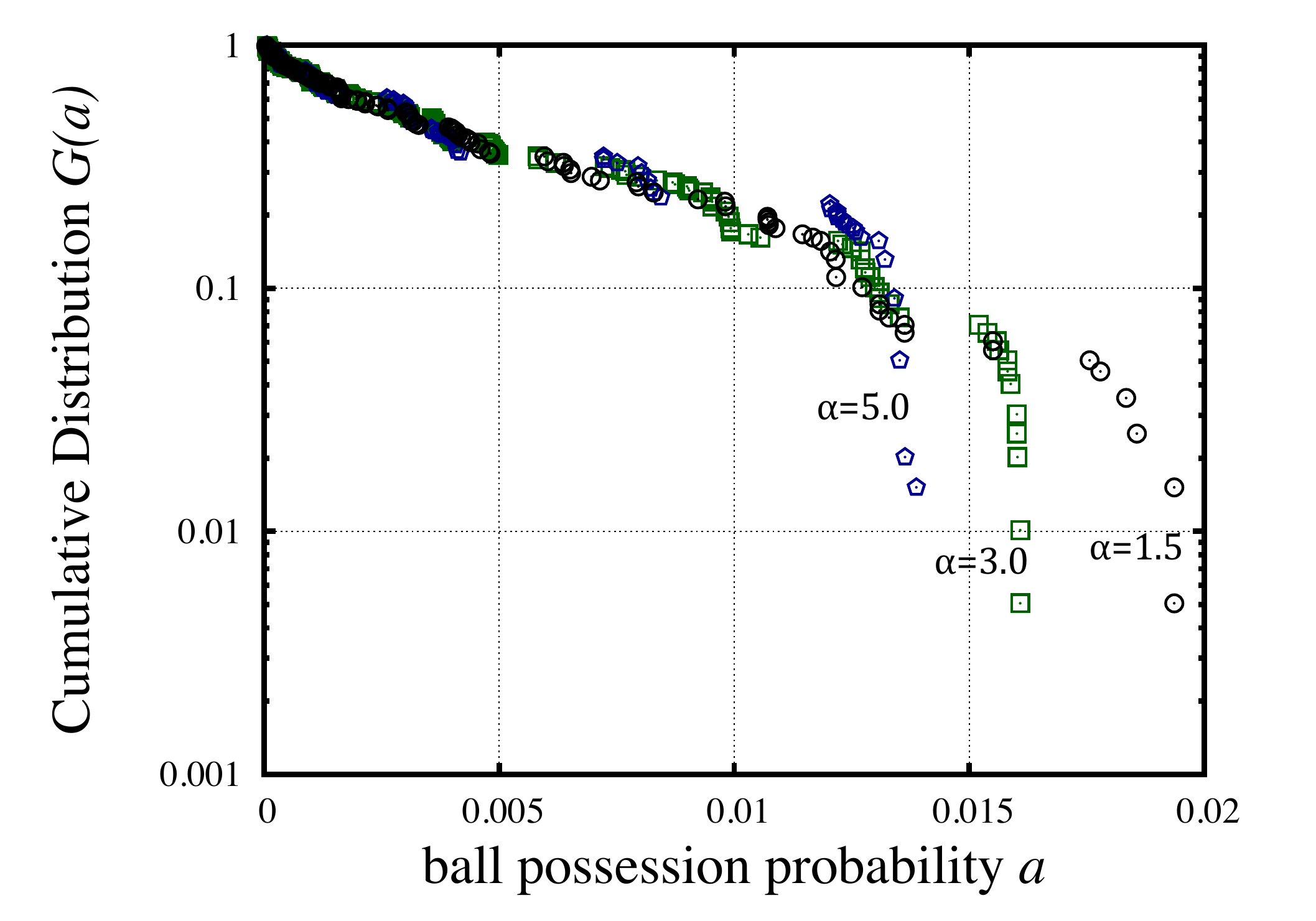}
		\caption*{(i)}
	\end{minipage}
	\begin{minipage}{.5\textwidth}
		\centering
		\includegraphics[width=7.5cm]{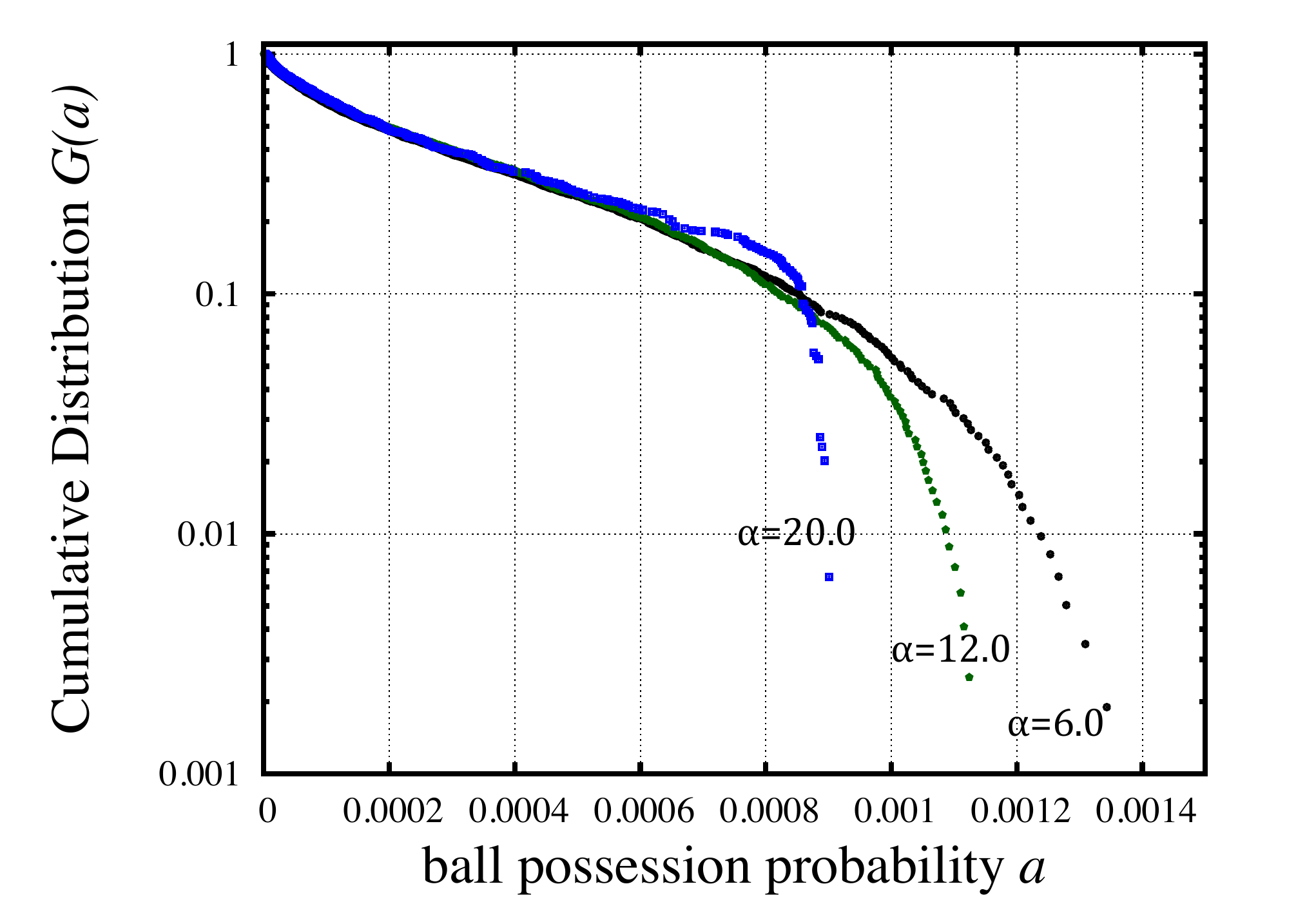}
		\caption*{(ii)}
	\end{minipage}
	\caption{Dependence of $G(a)$ on parameter $\alpha$: (i) $\beta=1.2$ and $\xi=1.0$ for $3\times6$, (ii) $\beta=0.3$, $\xi=4.0$ for $12\times24$.}
	\label{fig:CD-model-alpha}
\end{figure}

The dependence of $G(a)$ on $Q_{\alpha}(r_{ij})$ is shown in Fig. \ref{fig:CD-model-alpha}.
There exists a ball-possession probability 
above which $G(a)$ rapidly decreases, and it moves to the left when $\alpha$ increases.
However, the shape of $G(a)$ in the small region does not change largely. 
This property is independent of the field division. 

\begin{figure}[H]
	\begin{minipage}{.5\textwidth}
		\centering
		\includegraphics[width=7.5cm]{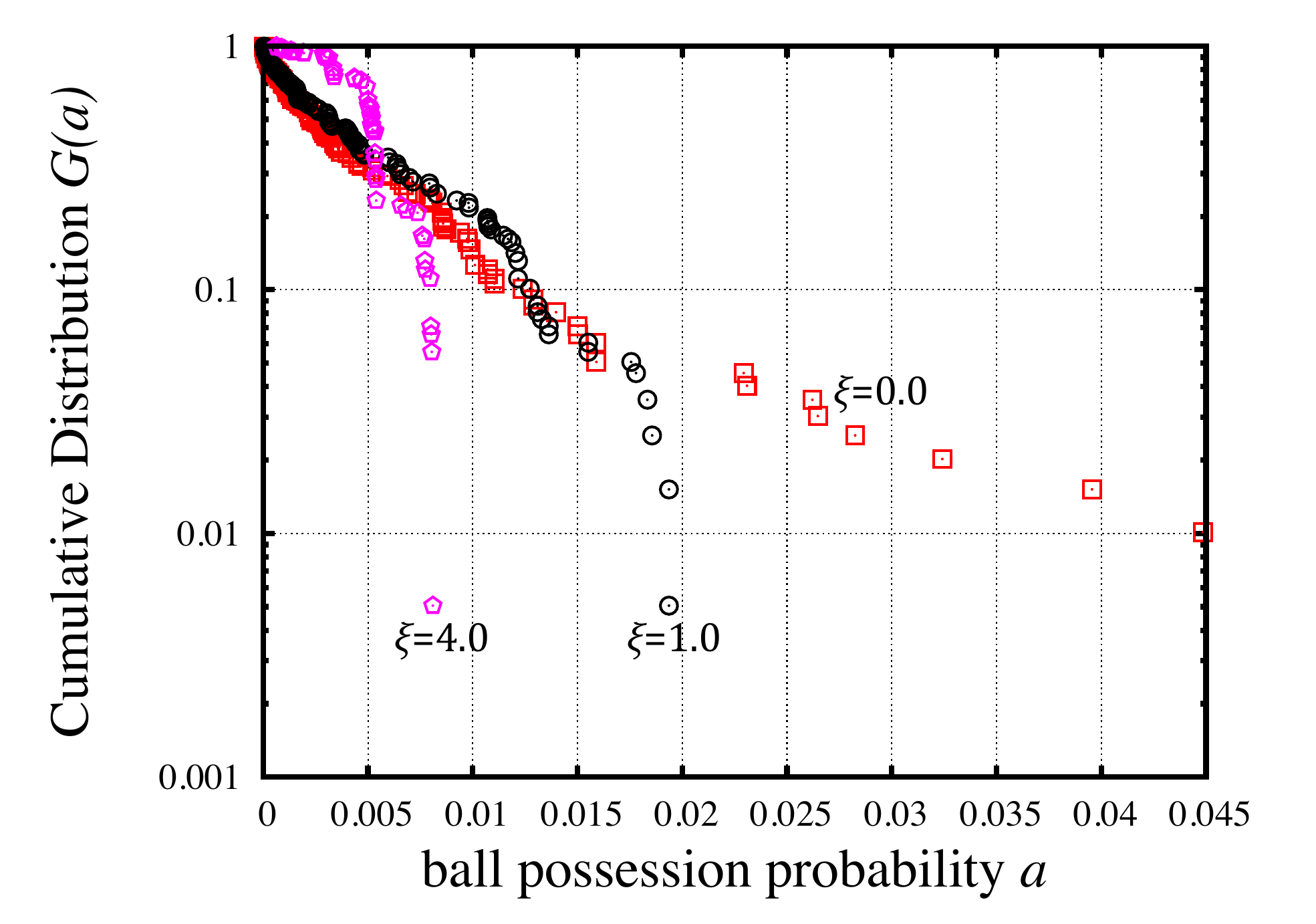}
		\caption*{(i)}
	\end{minipage}
	\begin{minipage}{.5\textwidth}
		\centering
		\includegraphics[width=7.5cm]{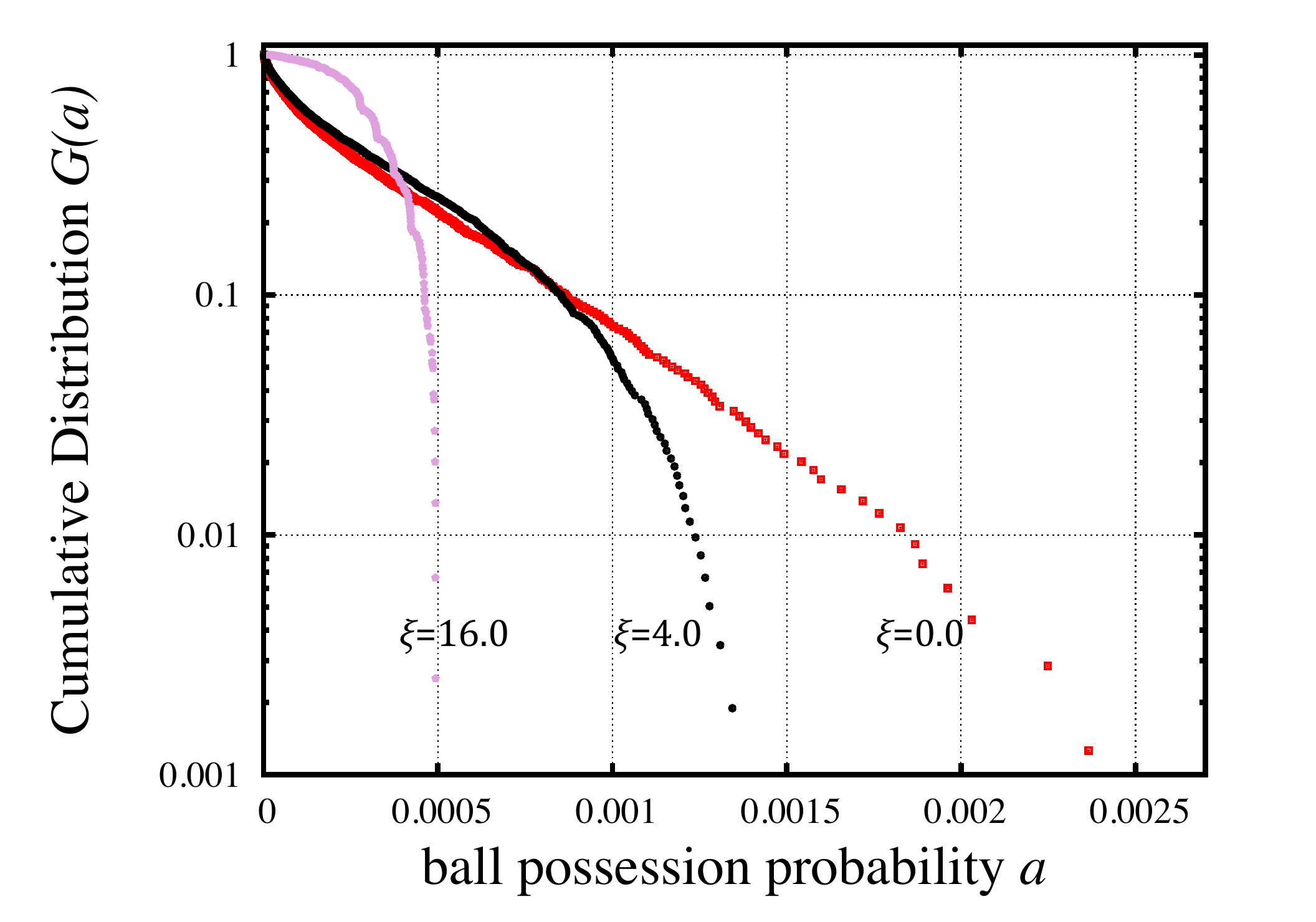}
		\caption*{(ii)}
	\end{minipage}
	\caption{Dependence of $G(a)$ on parameter $\xi$: (i) $\alpha=1.5$ and $\beta=1.2$ for $3\times6$, (ii) $\alpha=6.0$, $\beta=0.3$ for $12\times24$.}
	\label{fig:CD-model-l}
\end{figure}

\begin{figure}[H]
	\begin{minipage}{.5\textwidth}
		\centering
		\includegraphics[width=7.5cm]{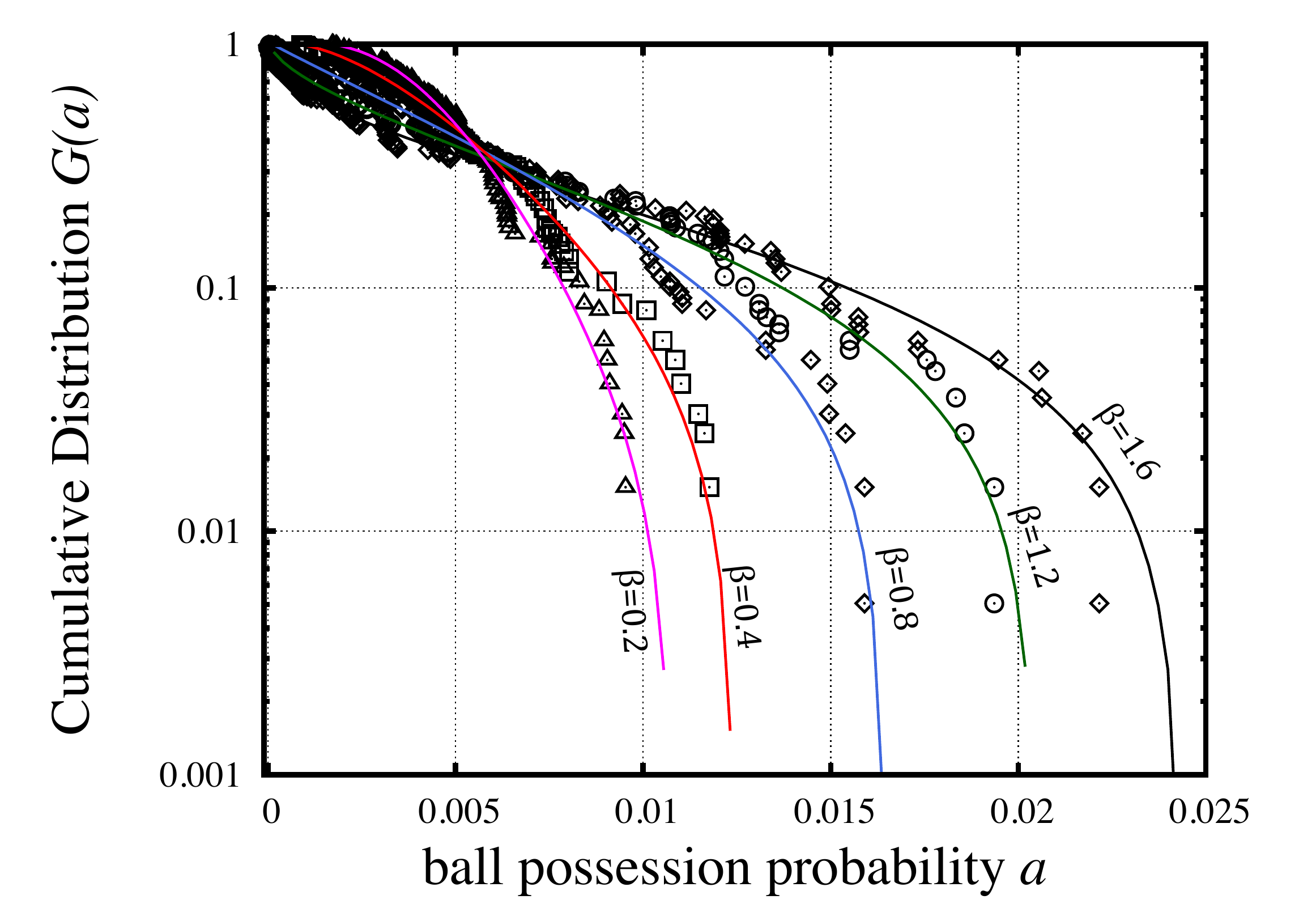}
		\caption*{(i)}
	\end{minipage}
	\begin{minipage}{.5\textwidth}
		\centering
		\includegraphics[width=7.5cm]{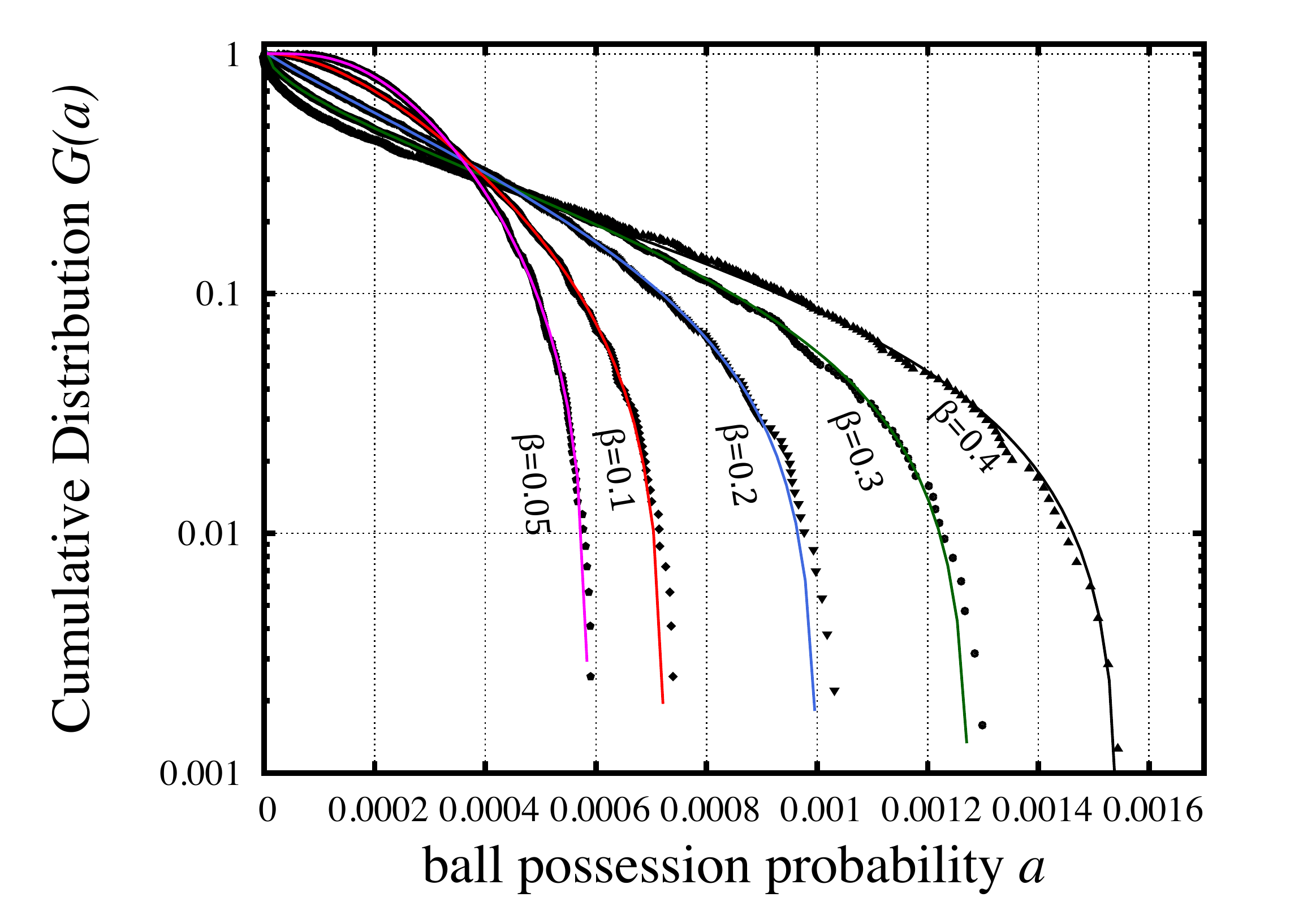}
		\caption*{(ii)}
	\end{minipage}
	\caption{Dependence of $G(a)$ on parameter $\beta$: (i) $\alpha=1.5$ and $\xi=1.0$ for $3\times6$, (ii) $\alpha=6.0$ and $\xi=4.0$ for $12\times24$.}
	\label{fig:CD-model-beta}
\end{figure}

The dependence of $G(a)$ on $R_{\beta,\xi}(L_{j})$ is shown in Figs. \ref{fig:CD-model-l} and \ref{fig:CD-model-beta}.
Especially for $\beta$-dependence shown in Fig. \ref{fig:CD-model-beta},
it is found that $G(a)$ is fitted well
by the truncated gamma distribution, and $\nu$ decreases monotonically against $\beta$ as shown in Tables \ref{tab:fitting-model} and \ref{tab:fitting-model2}.
Since $\nu$ obtained from the real data was about 0.3 for each networks, we can reproduce this value by choosing $\beta$ appropriately in the model.

\begin{table}[H]
  \centering
  \caption{The parameters for fittings of $G(a)$ with $\alpha=1.5$, $\xi=1.0$. $\beta$ is controll parameter ($3\times6$). }
  \vspace*{-0.3cm}
    \begin{tabular}{ccccccc}
    \toprule
$\beta$&  & $\nu$   &  $\lambda$            & $a_{\rm max}$         & $a_{\rm min}$           &   \\  \toprule
0.2       &  & 5.37     & $9.6\times10^{-4}$ & $1.1\times10^{-2}$  & $1.7\times10^{-3}$  &    \\
0.4       &  & 2.45     & $2.2\times10^{-3}$ & $1.2\times10^{-2}$  & $8.6\times10^{-4}$  &     \\
0.8       &  & 0.93     & $7.1\times10^{-3}$ & $1.6\times10^{-2}$  & $1.3\times10^{-4}$  &      \\
1.2       &  & 0.47     & $1.9\times10^{-2}$ & $2.0\times10^{-2}$  & $3.9\times10^{-5}$    &        \\
1.6       &  & 0.30     & $4.1\times10^{-2}$  & $2.4\times10^{-2}$  & $8.6\times10^{-6}$    &         \\
    \bottomrule
    \end{tabular}%
  \label{tab:fitting-model}%
\end{table}%

\begin{table}[H]
  \centering
  \caption{The parameters for fittings of $G(a)$ with $\alpha=6.0$, $\xi=4.0$. $\beta$ is controll parameter ($12\times24$).}
  \vspace*{-0.3cm}
    \begin{tabular}{ccccccc}
    \toprule
$\beta$ &  & $\nu$   & $\lambda$          & $a_{\rm max}$             & $a_{\rm min}$            &      \\ \toprule
0.05      &  & 4.36  & $8.0\times10^{-5}$ & $5.9\times10^{-4}$      & $6.1\times10^{-5}$     &      \\
0.1        &  & 2.24  & $1.6\times10^{-4}$ & $7.3\times10^{-4}$      & $2.2\times10^{-5}$     &       \\
0.2        &  & 0.85  & $5.0\times10^{-4}$ & $1.0\times10^{-3}$      & $8.3\times10^{-6}$     &        \\
0.3        &  & 0.47  & $1.1\times10^{-3}$ & $1.3\times10^{-3}$      & $2.2\times10^{-6}$     &        \\
0.4        &  & 0.37  & $1.5\times10^{-3}$ & $1.6\times10^{-3}$      & $1.2\times10^{-7}$     &        \\    \bottomrule
    \end{tabular}%
  \label{tab:fitting-model2}%
\end{table}%

\clearpage

\clearpage

\subsection{Networks based on a transitive matrix}
By determining $Q_{\alpha}(r_{ij})$ and $R_{\beta,\xi}(L_{j})$,
the ball-passing probabilities between all pairs of nodes are obtained.
Using this probability, we can make a ball-passing sequence.
Actually we created networks in the way that all pairs of ball-passing and receiving nodes were connected by an undirected edge.
Here, we focus on the networks of Spain, Holland, and North Korea in the real data.
We created networks with the same $M$ as these real networks.

It is found that the values of $\ell$ and $C$ are in good agreement with those obtained from real networks where $\alpha$, $\beta$, and $\xi$ are given as shown in Table \ref{tab:character-model}.
This result suggests that the Markov-chain model can reproduce statistical properties of real data.
\\
\begin{table}[H]
  \centering
  \caption{The values characterizing each network obtained from the numerical simulation.}
    \vspace*{-0.3cm}
    {\footnotesize 
    \begin{tabular}{ccccccccccc}
    \toprule
    \multicolumn{4}{l}{Real Networks}                      &       &     \multicolumn{6}{l}{Numerical Results} \\ \cmidrule{1-4}\cmidrule{6-11}
    Team            & $M$& $\ell$ & $C$   &   & $\alpha$ &$\beta$ & $\xi$ & &$\ell$                   & $C$                  \\ \toprule
    Spain            & 688  & 3.11 & 0.29  &   & 1.0          & 1.8       & 1.0 & &$3.24\pm0.07$ & $0.28\pm0.03$ \\
    Holland         & 501  & 3.05 & 0.23  &   & 1.5          & 1.7       & 1.0 & &$2.91\pm0.07$ & $0.23\pm0.03$ \\
    North Korea  & 272  & 3.49 & 0.11  &   & 1.5          & 1.0       & 1.0 & &$3.49\pm0.15$ & $0.11\pm0.03$ \\
    \bottomrule
    \end{tabular}%
    }
  \label{tab:character-model}%
\end{table}%

\clearpage

\section{Discussion}
The main result of our numerical simulation is that 
$G(a)$ obtained from our model can be fitted with the truncated gamma distribution as in the case of the degree distribution from the real data.
In addition, we have also found that a network created from our model
has similar structural properties to the real data.
Judging from these results, we conclude that our model incorporates essential features of real football games.

\begin{figure}[H]
\centering
	{\includegraphics[width=8cm]{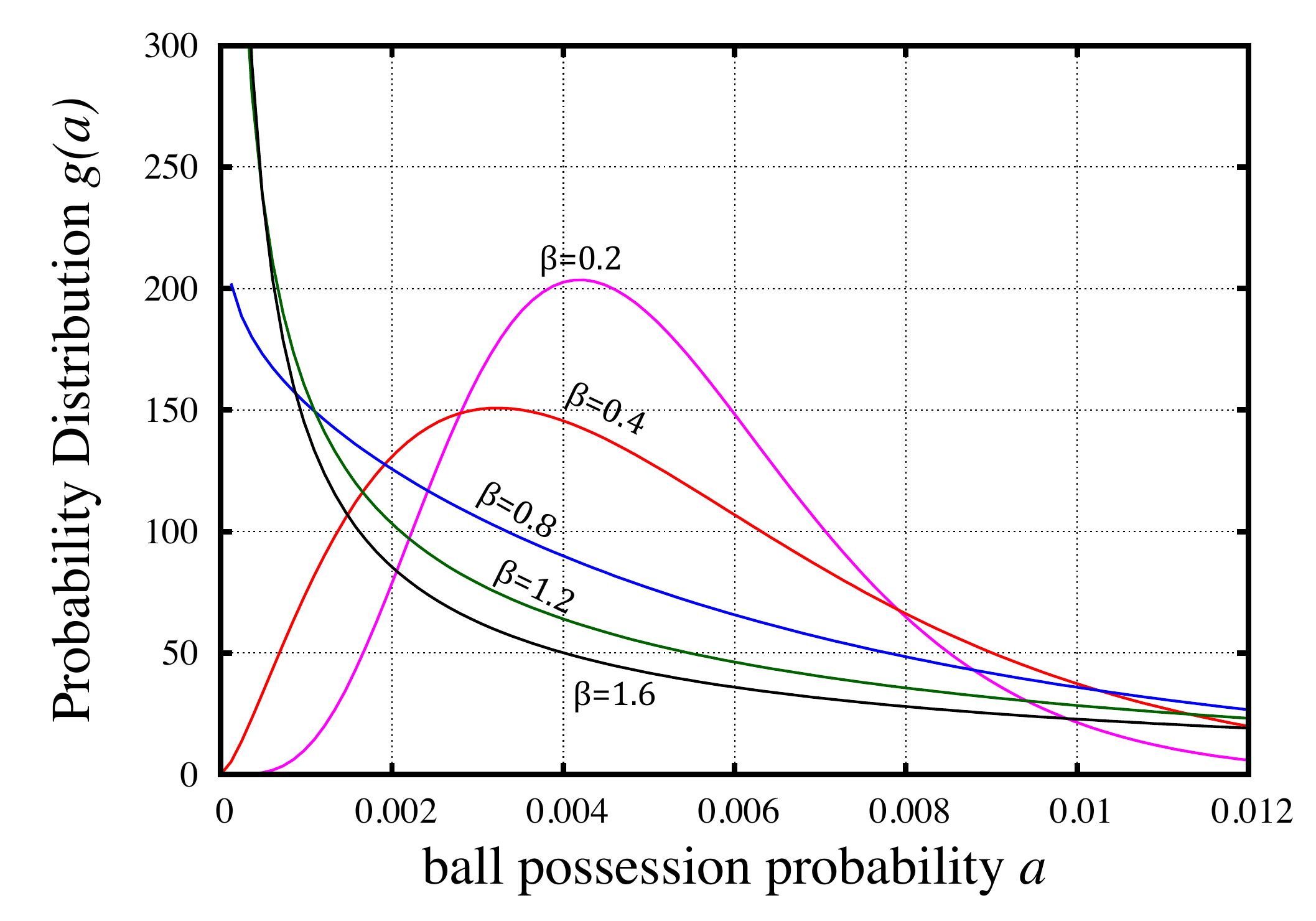}}
\vskip -\lastskip
	\caption{Change of the shape of $g(a)$ with $\beta$}
	\label{fig:CD-model-beta2}
\end{figure}

We have also obtained the result that $\nu$ decreases monotonically against $\beta$. 
The shape of $G(a)$ depends mainly on $R_{\beta,\xi}(L_{j})$, and the result can be explained qualitatively as follows.
When $\beta$ is small, passes are made almost at random because the existence probability $R_{\beta,\xi}(L_{j})$ of each node has almost the same value.
Then, the values $a_{i}^{(t)}$ of all nodes also become almost the same,
and the probability distribution $g(a)$, defined as $g(a)\equiv -dG(a)/da$, has a peak as shown in Fig. \ref{fig:CD-model-beta2}.
This brings about the truncated gamma distribution with large $\nu$.  
However, as $\beta$ increases, areas where each player can move become limited.
Namely, the number of nodes having small existence probability increases, 
and the peak of the distribution moves to the left as shown in Fig. \ref{fig:CD-model-beta2}.
Such a change of the distribution corresponds to the decreasing of the value of $\nu$ in the truncated gamma distribution.
Thus, the value of $\nu$ becomes small if $\beta$ becomes large.
It is noted that $g(a)$ without the peak
corresponds to the truncated gamma distribution with $\nu$ smaller than 1.
We can also explain the reason why $G(a)$ rapidly decreases where $a$ is large.
In section 4, we have found that this behavior appears where $\xi$ is positive.
In such a condition, $R_{\beta,\xi}(L_{j})$ is constant in the part of low $L_{j}$, and $a_{i}^{(t)}$ is less likely to be large.
In other words, the cutoff appeared because of the existence of threshold of the ball possession probability.

\begin{figure}[H]
\centering
	{\includegraphics[width=8cm]{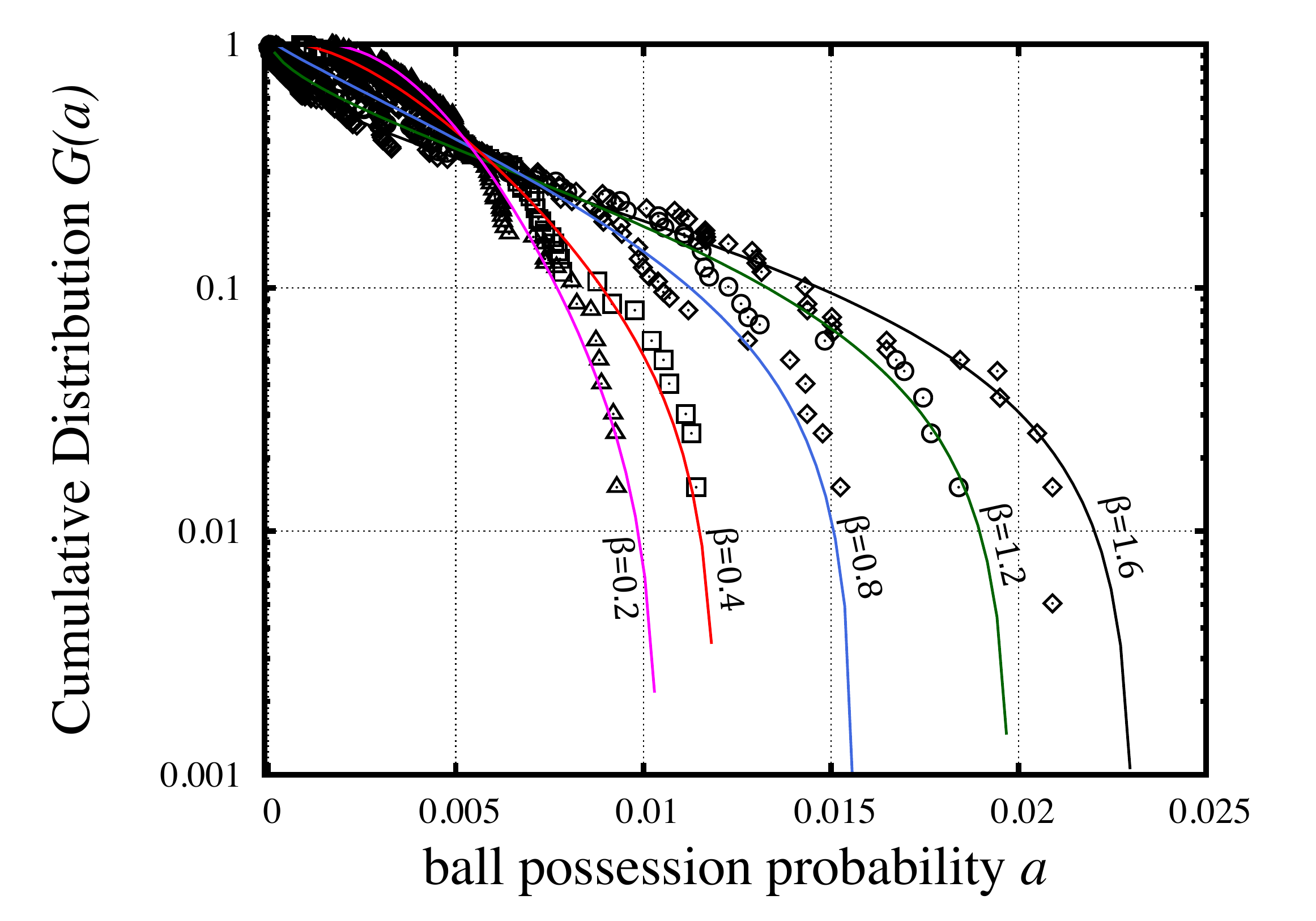}}
\vskip -\lastskip
	\caption{Dependence of $G(a)$ on parameter $\beta$ where $g=0.5$, $\alpha=1.5$ and $\xi=1.0$.}
	\label{fig:CD-model-beta-opp}
\end{figure}

 For simplicity, we have ignored the existence of the opponent team in this paper.
 However,  the passing sequence of the real game is interrupted by opponent team, and it is the superposition of some passing sequences beginning with different initial nodes.
 We can easily take this effect into account. 
 First, we add one node which represents opponent player to the model, and express it as ``opp".
Second, we change the transitive probability as follows.
 (1) If the two nodes $i$ and $j$ represent the same players, then $P_{i\to j}=0$.
 (2) Otherwise, $P_{i\to j}$ is defined as:
 \begin{equation}
 \label{eq:opp}
	\begin{cases}
		P_{i\to j}\propto (1-g) \times Q_{\alpha}(r_{ij})\times R_{\beta,\xi}(L_{j})\\
		P_{i\to \rm opp}\propto g \\
		P_{{\rm opp}\to j}\propto R_{\beta,\xi}(L_{j}) \\
		P_{{\rm opp}\to {\rm opp}} = 0, \\
	\end{cases}\\ 
\end{equation}
 where  $g$ represents the probability of interruption by a opponent player.
 Figure \ref{fig:CD-model-beta-opp} shows the dependence of $G(a)$ on parameter $\beta$ where $g=0.5$, $\alpha=1.5$ and $\xi=1.0$. 
It is found that $G(a)$ is fitted well by the truncated gamma distribution.
Namely, we can obtain the same distribution even if there are the interruption of passing sequences.

The factor $R_{\beta,\xi}(L_{j})$ determines the moving distance of each player.
In the present study, we have assumed that $R_{\beta,\xi}(L_{j})$ is an exponential function.
Other decreasing functions, e.g. linear, Gaussian, or power-law functions are also considered.
Actually, we confirmed that $G(a)$ can be fitted by the truncated gamma distribution when $R_{\beta,\xi}(L_{j})$ is a Gaussian.
However, we cannot reproduce the truncated gamma distribution when $R_{\beta,\xi}(L_{j})$ is the linear function and power-function.    
Detailed examination for the dependence of $G(a)$ on $R_{\beta,\xi}(L_{j})$ is needed.

In section 4, we have also examined the effect of the field division.
$G(a)$ follows the truncated gamma distribution independent of the field division.
In the real data analysis, more detailed data may be obtained if we divide the field more finely.
However, the number of passes is constant in one game, and the degree of each node becomes small when the division is more finely.
Hence, it is difficult to discuss statistical properties of such networks.
For our network analysis, $3\times6$ division seems to be suitable way of dividing the soccer field. 

\begin{figure}[H]
	\begin{minipage}{.5\textwidth}
		\centering
		\includegraphics[width=9cm]{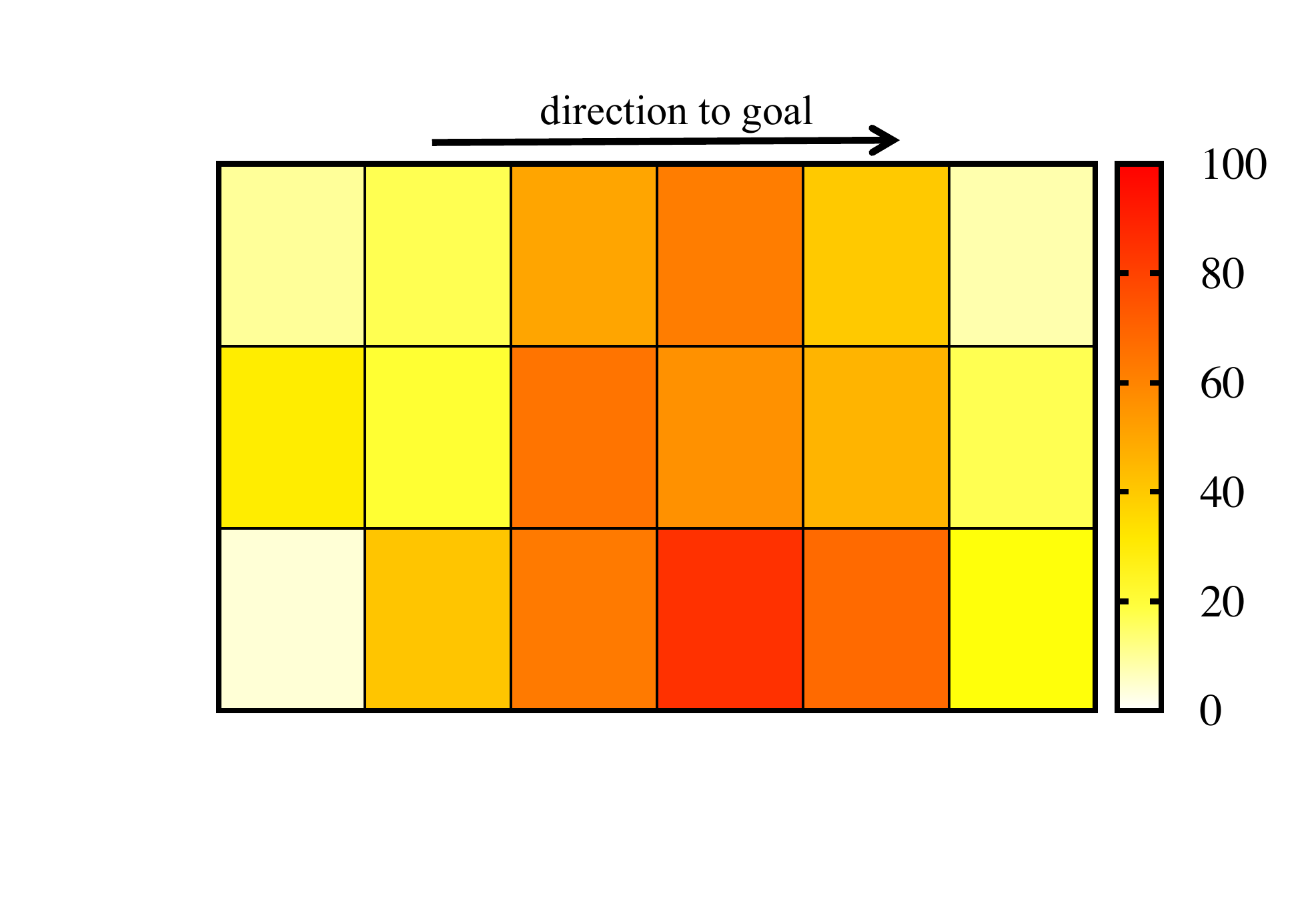}
		\caption*{(a) Japan}
	\end{minipage}
		\begin{minipage}{.5\textwidth}
			\centering
			\includegraphics[width=9cm]{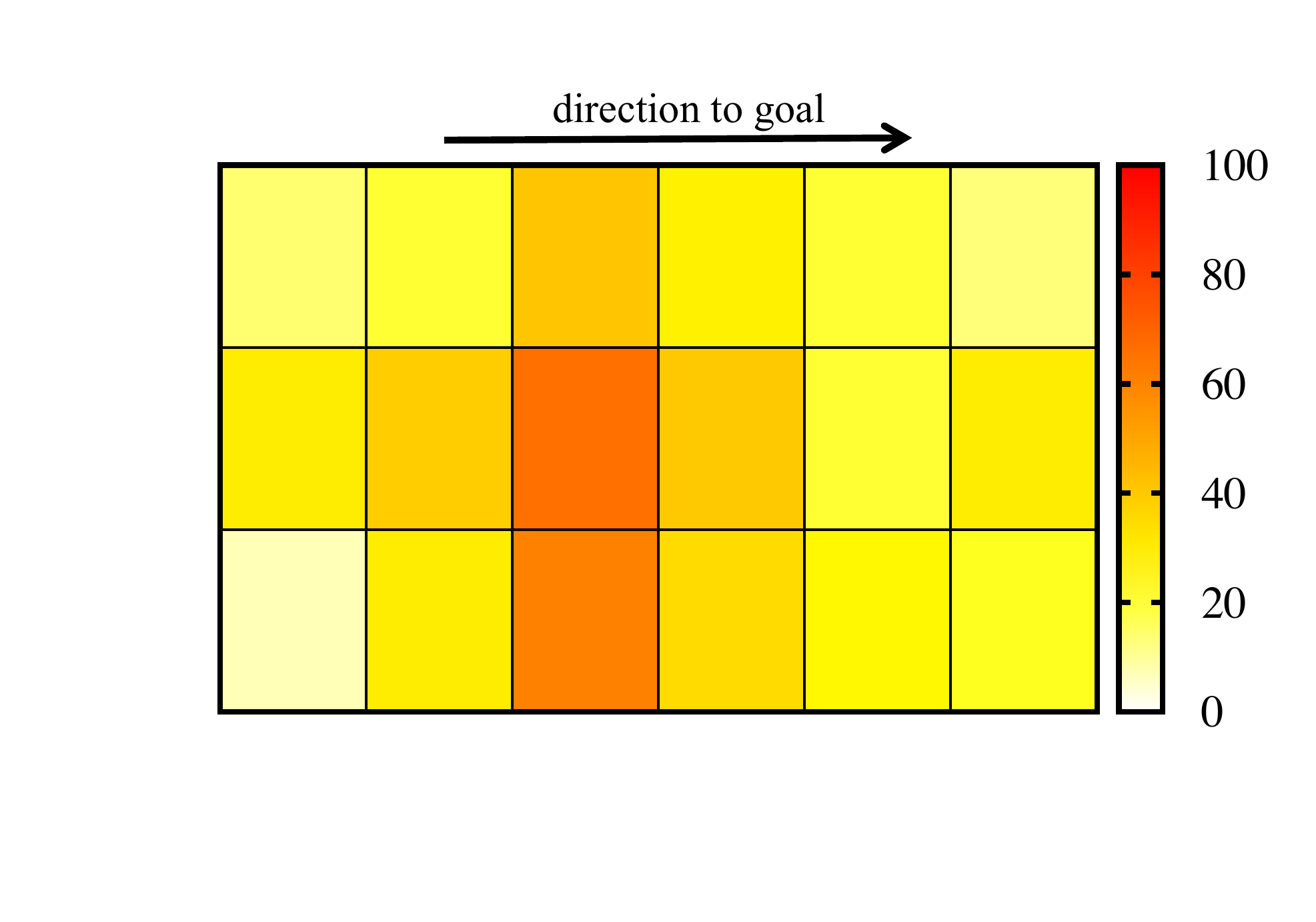}
			\caption*{(b) North Korea}
	\end{minipage}
	\caption{Heat maps obtained from the game (iii). The gradation of shading shows the sum of degrees of all nodes on one area. }
	\label{fig:heatmap}
\end{figure}

Our findings indicate that the degree distributions in the real data reflect the motion of each player which is characterized by $R_{\beta,\xi}(L_{j})$.
If it is approximated by the exponential function such as Eq. \eqref{eq:R}, we can define the characteristic moving distance $L_{\rm relax}$ as $R_{\beta,l}(L_{\rm relax})=1/e$.
From Eq. \eqref{eq:R}, it is expressed as $L_{\rm relax}=\xi+1/\beta$.
Since we have obtained $\nu = 0.34\pm 0.07$ for each teams in the real data, $L_{\rm relax}$ is calculated as $L_{\rm relax}=1.625$ for $3\times6$ and $L_{\rm relax}=6.5$ for $12\times24$ from the values of $\beta$ and $\xi$ in Tables \ref{tab:fitting-model} and \ref{tab:fitting-model2}.
Therefore, the distance in which each player can move is very limited.
In football games, each team tries to move the ball to the opponent goal in an efficient way. 
Then, each player moves around only his own home position, and passes the ball.
Namely, this result shows the characteristics of football that each player has his own role corresponding to their home position.
In the analysis of the real data, we have also examined the edge-multiplicity distribution, and found that almost $70\%$ of edges have multiplicity one, and there are a few high-multiple edges.
The low-multiple edges seem to correspond to passes which are made at random.
On the other hand, the high-multiple edges are considered to correspond to the routine passes which are related to the strategies of each team.
Figure \ref{fig:heatmap} shows the heat maps obtained from the game (iii). 
The gradation of shading shows the sum of degrees of all nodes on one area.
The heat map gives some information about strategies of each team. 
For example, Japan in the game (iii) uses side areas many times, and it is considered that this team often uses the side attack.
Then, edges on the side areas have high multiplicity in this case.
The heat maps also reflect the activity of each team.
Actually, Japan in this game mainly makes passes in the areas near the opponent-team goal, while North Korea makes a few passes in the areas near the own-team goal.
Then it shows that Japan dominated the game.

Global properties, such as degree distribution and edge-multiplicity distribution, of the ball-passing network are similar in different teams and games, but local properties are different.
In fact, the degree distribution of each team follows the same distribution, and the edge-multiplicity distribution decreases almost exponentially, while the heat maps of each team are different.
In this paper, we have mainly focused on the global properties of networks which do not depend on the details of the game.
And, the local properties which are related to the strategies of each team are also interesting.

We comment on relation to the previous studies.
Yamamoto and Yokoyama have reported that ball passing forms a small-world network, and its degree distribution is described by the power-law, power-law with exponential cutoff, or the exponential distributions \cite{Yamamoto2010,Yamamoto2011b}.  
Our position-dependent network is a natural extension of the networks in the previous ones.
Moreover, our analysis reveals that the degree distribution can be characterized by using the single distribution, truncated gamma distribution, while several distributions were used for fitting in the previous study.

Finally, some future works are proposed.
We can readily obtain directed networks by considering the directions of passes.
The directed network is expected to represent the defense and offense in football games. 
For modeling based on Markov chain, a more realistic model can be considered.
For example, since each player mainly moves along longitudinal direction in real football games, some anisotropic effects should be introduced in $Q_{\alpha}(r_{ij})$ and  $R_{\beta,\xi}(L_{j})$.
Moreover, the method of creating networks we proposed in this paper can be applied to other ball games such as basketball, rugby, and handball.
From the viewpoint of network analysis, our model is associated with the spatial network \cite{Barthelemy2011}, in which
the nodes are distributed in a $D$-dimensional space with density $\rho (x)$, and two nodes are connected by an edge whenever their distance is less than a given threshold value.
Our model resemble the spatial-network model, in that two nodes $i$ and $j$ can make a pass if their distance $r_{ij}$ is less than $\alpha$ (see Eq. \eqref{eq:Q} of $Q_{\alpha}(r_{ij})$ for reference); $\alpha$ works as the threshold.
The spatial-network model is deterministic, but our model involves a stochastic rule.
We expect that statistical properties of the ball-passing network are analytically studied as ``stochastic spatial network".

\clearpage

\section{Conculusion}
We have created the position-dependent network of ball passing in football games in this paper. 
Each network has short path length $\ell \approx 3.3$ which is close to $\ell_{\rm rand} \approx 4.4$, and high clustering coefficient $C \gg C_{\rm rand}$.
Degree distribution is fitted well by the truncated gamma distribution.
In the Markov-chain model, we define the transition probability $P_{i\to j}$ as the product of the two factors for the distance of passes $Q_{\alpha}(r_{ij})$ and for the existence probability of the player receiving a pass $R_{\beta,\xi}(L_{j})$.
The ball-possession probability distribution reproduces the truncated gamma distribution.
Also, it is found that the network created by our model is similar to those from the real data.

\section{Acknowledgements}
The present work was supported through the Takuetsu Program, by the Ministry of Education, Culture, Sports, Science and Technology.  
\clearpage

\bibliography{./reference} 
\clearpage

\section*{List of Symbols}

\renewcommand{\arraystretch}{1.5}
\begin{table}[H]
	\centering
	\vspace*{-0.3cm} 
	\label{tb:symbol}
 		\begin{tabular}{ll}
		\toprule
		 $N$              & the total number of nodes \\
		 $M$              & the total number of edges \\
		 $\ell$            &  mean path length \\
		 $C$               &  clustering  coefficient \\
		 $k$                &  degree \\
		 $f(k)$             &  probability distribution of degree \\
		 $F(k)$            &  cumulative distribution of degree \\
		 $\nu$             & shape parameter of the truncated gamma distribution \\
		 $\lambda$      & scale parameter of the  truncated gamma distribution \\
		 $a_{i}^{(t)}$     & ball-possession probability for the node $i$ at time $t$ \\
		 $\bm{a}^{(t)}$  & probability vector of $a_{i}^{(t)}$\\
		 $\tau$             & time steps until $\bm{a}^{(t)}$ reaches the steady state\\
		  $g(a)$            & probability distribution of $\bm{a}^{(\tau)}$\\
		 $G(a)$            &  cumulative distribution of $\bm{a}^{(\tau)}$\\
		 $P_{i\to j}$       & ball-passing probability from the node $i$ to the node $j$  \\
		 $\bm{P}$         & transitive matrix\\
		 $r_{ij}$             & the distance between the two nodes $i$ and $j$ \\
		 $L_{j}$               &  the distance of the node $j$ from its home position\\
		 $Q_{\alpha}(r_{ij})$ & the factor for the distance of passes in $P_{i\to j}$\\
		 $R_{\beta,\xi}(L_{j})$ & the factor for the existence probability of the player receiving a pass \\
		 $\alpha$              & the parameter in $Q_{\alpha}(r_{ij})$ \\
		 $\beta$               & the parameter in  $R_{\beta,\xi}(L_{j})$\\
		 $\xi$                & the parameter in  $R_{\beta,\xi}(L_{j})$\\
		 $L_{\rm relax}$ & the characteristic moving distance of each player\\
		 $g$                   & the ball-passing probability to the opponent team \\
		\bottomrule 
		\end{tabular}
\end{table}
\renewcommand{\arraystretch}{1.0}

\clearpage

\end{document}